\documentclass[aps,prb,superscriptaddress,showpacs,floatfix,twocolumn]{revtex4-1}
\usepackage{graphicx,amsmath,amssymb,xspace,epsfig,float,multirow,subfigure,tabularx}

\pdfoutput=1

\begin{document}

\title{Chiral universality class of the normal-superconducting and the
exciton condensation transition on the surface of topological insulator }
\author{Dingping Li}
\email{lidp@pku.edu.cn}
\affiliation{School of Physics, Peking University, Beijing
100871, \textit{China}} \affiliation{Collaborative Innovation Center of
Quantum Matter, Beijing, China}
\author{Baruch Rosenstein}
\email{vortexbar@yahoo.com,correspondent author}
\affiliation{Electrophysics Department, National Chiao Tung University, Hsinchu 30050, \textit{Taiwan,
R. O. C}} \affiliation{Physics Department, Ariel University, Ariel 40700, Israel}
\author{B.Ya. Shapiro}
\email{shapib@mail.biu.ac.il}
\affiliation{Physics Department, Bar-Ilan University, 52900 Ramat-Gan, Israel}
\author{I. Shapiro}
\affiliation{Physics Department, Bar-Ilan University, 52900 Ramat-Gan, Israel}
\date{\today }

\begin{abstract}
New two dimensional systems like surface of topological insulator and
graphene offer a possibility to experimentally investigate situations
considered "exotic" just a decade ago. One of those is the quantum phase
transition of the "chiral" type in electronic systems with relativistic
spectrum. Phonon mediated ("conventional") pairing in the Dirac semimetal
appearing on the surface of topological insulator leads to transition into a
chiral superconducting state, while exciton condensation in these gapless
systems has been envisioned long time ago in the physics of the narrow band
semiconductors. Starting from the microscopic Dirac Hamiltonian with local
attraction or repulsion, the BCS type gaussian approximation is developed in
the framework of functional integrals. It is shown that due to an
"ultra-relativistic" dispersion relation there is a quantum critical point
governing the zero temperature transition to a superconducting or the
exciton condensed state. The quantum transitions\ that have critical
exponents very different from the conventional ones. They belong to the
chiral universality class. We discuss the application of these results to
recent experiments in which surface superconductivity was found in
topological insulators and estimate feasibility of the phonon pairing.
\end{abstract}

\maketitle

\section{Introduction}

Topological insulator (TI) is a novel state of matter in materials with
strong spin - orbit interactions that create topologically protected surface
states \cite{Zhang}. The electrons (holes) in these states have a linear
dispersion relation, see Fig. 1, and can be described approximately by a
(pseudo) relativistic two dimensional (2D) Hamiltonian. The system realizes
an "ultra-relativistic" 2D electron or hole conducting liquid along with
much better studied graphene\cite{Katsnelson}, a 2D one layers sheet of
carbon atoms that became a paradigms example of the Dirac semi-metal. In the
context of graphene certain quantum phase transitions were theoretically
contemplated. The superconductivity in graphene has been repeatedly
considered\cite{supergraphene}, however despite great experimental efforts
was never achieved. The 2D dimensional phonons seem to be unable to overcome
strong Coulomb repulsion in order to create a Cooper pair. The same can be
said about attempts to achieve exciton condensate in graphene that was
proposed\cite{Khveshchenko,Gamayun} even before its discovery. Apparently
the repulsion is not strong enough either to create stable electron - hole
bound states\cite{graphenechiral}. The surface of topological insulator
therefore became a prime candidate to realize the quantum transitions.

It is known for a long time that similar 2D and quasi-2D metallic systems
like the surface metal on twin planes \cite{Shapiro}, layered materials
(strongly anisotropic high $T_{c}$ cuprates\cite{Wen} or organic
superconductors\cite{organic}) may develop 2D (surface) superconductivity.
This phenomenon became known as "localized superconductivity"\cite{Buzdin}.
Since best studied TIs possess a quite standard phonon spectrum \cite%
{phononexp}, it was predicted recently \cite{DasSarma,Li14} that they become
superconducting. The predicted critical temperature of order of $1K$ is
rather low (despite a fortunate suppression of the Coulomb repulsion due to
a large dielectric constant $\varepsilon \sim 50$), the nature of the
"normal" state (so-called 2D Weyl semi-metal) might make the superconducting
properties of the system unusual. The ultra-relativistic nature manifests
itself mostly when the Weyl cone is very close to the Fermi surface.
Especially interesting is the case (that actually was originally predicted
for the [111] surface of $Bi_{2}Te_{3}$ and $Bi_{2}Se_{3}$\cite{Zhang1})
when the chemical potential coincides with the Dirac point. Although
subsequent ARPES experiments\cite{Zhang} show the location of the cone of
surface states order tenths of $eV$ off the Fermi surface; there are
experimental means to shift the chemical potential, for example by the bias
voltage \cite{bias}.

Unlike the more customary poor 2D metals with several small pockets of
electrons/holes on the Fermi surface (in semiconductor systems or even some
high $T_{c}$ materials\cite{Wen}), the electron gas TI has two peculiarities
especially important when pairing is contemplated. The first is the bipolar
nature of the Dirac spectrum: there is no energy gap between the upper and
lower cones. The second is that \ the spin degree of freedom is a major
player in the quasiparticle dynamics. This degree of freedom determines the
pairing channel. The pairing channel problem was studied theoretically on
the level of the Bogoliubov-deGennes equation \cite{Herbut}. Both $s$-wave
and $p$-wave are possible and compete due to the breaking of the bulk
inversion symmetry by the surface. Various pairing interactions were
considered to calculate the DOS measured in $Cu_{x}Bi_{2}Se_{3}$ using
self-consistent analysis \cite{Sato}. As mentioned above the most intriguing
case is that of the small chemical potential that has not been addressed
microscopically. It turns out that it is governed by{\LARGE \ }a quantum
critical point (QCP)\cite{Sachdev}.

The concept of QCP at zero temperature and varying doping constitutes a very
useful language for describing the microscopic origin of superconductivity
in high $T_{c}$ cuprates and other "unconventional" superconductors\cite{Wen}%
. Superconducting transitions generally belong to the $U\left( 1\right) $
class of second order phase transitions\cite{Herbutbook}, however it was
pointed out a long time ago\cite{Rosenstein} that, if the normal state
dispersion relation is "ultra-relativistic", the transition at zero
temperature as function of parameters like the pairing interaction strength
is qualitatively distinct and belongs to chiral universality classes
classified in ref. \cite{Gat}. The term "chiral" appears following the
corresponding discussion of the well studied both theoretically and
experimentally chiral symmetry breaking transition in Quantum
Chromodynamics. Attempts to experimentally identify second order transitions
governed by QCP in condensed matter included quantum magnets \cite{Sachdev},
superconductor - insulator transitions\cite{SCinsulator} and more recently
exciton condensate in graphene\cite{Katsnelson,CastroNeto} and other Dirac
semi - metals including TI. In the last two cases the broken symmetry is
also often termed "chiral".

Exciton condensation is a very old concept in low dimensional narrow gap
semiconductor physics. The best-studied exciton condensate is the quantum
Hall bilayer at half-filled Landau levels\cite{QHE}. Here ingenious methods
had been developed to separately contact the two layers so that one can
directly probe the order parameter via counterflow superfluidity along the
layers and tunneling between the layers. The same idea was extended to
bilayer graphene and recently to TI\cite{excitonTI}.

In addition Dirac semimetal was realized in cold atom system\cite{cold}
(following the realization in 2D known as the "synthetic graphene").
Interestingly the sign and strength of the interaction can be controlled.
The Dirac semimetal in optically trapped cold atom systems \cite{cold} is
well suited to study this fascinating phenomenon. The Dirac semimetal in
optically trapped cold atoms\cite{cold} offers a well controllable system in
which this phenomenon occurs both for repulsive interaction (chiral symmetry
breaking) and the attractive one (superconductivity).

In this paper the quantum phase transitions in Dirac semimetal due to local
interactions both attractive (superconductivity) and repulsive (exciton
condensation) are studied with emphasis on their distinct criticality. The
critical exponents belong to chiral universality classes that are identified.

In Section II the general framework that allows to study the surface
superconductivity and exciton condensation in general Dirac semi-metal (TI,
not necessarily time reversal and reflection invariant or some other of the
numerous systems being identified recently) with a general local interaction
is presented. The local coupling strength $g$, chemical potential $\mu $ and
temperature $T$ will be kept general ($g$ is negative for repulsion leading
to the exciton condensation or positive leading to superconductivity). Since
the symmetry analysis is crucial, we first discuss the space and spin
rotations. In Section III we concentrate on the simplest time reversal and
reflection invariant Dirac model and identify its spontaneous symmetry
breaking patterns. In Section IV the phase diagram for $g>0$ is obtained for
arbitrary temperature $T$ and chemical potential much smaller than the Debye
energy $T_{D}$. The latter condition is the main difference from the
conventional BCS model in which $\mu >>T_{D}$. A quantum critical point at $%
T=\mu =0$ when the coupling strength $g$ reaches a critical value $g_{c}$
dependent on the cutoff parameter $T_{D}$. We concentrate on properties of
the superconducting state in a part of the phase diagram that is dominated
by the QCP. Various critical exponents are obtained. In particular, the
coupling strength dependence of the coherence length is $\xi \propto \left(
g-g_{c}\right) ^{-\nu }$ with $\nu =1$ , the order parameter scales as $%
\Delta \propto \left( g-g_{c}\right) ^{\beta }$, $\beta =1$. For the
repulsion similar transition occurs in the exciton channel in Section V. The
critical exponents beyond mean field and experimental feasibility of
superconductivity are discussed in Section VI.

\section{Generalized mean field approximation for local four - Fermi
interactions}

\subsection{Hamiltonian and the partition function}

We consider the second quantized electron Hamiltonian via four-Fermi local
coupling of strength $g$
\begin{gather}
H=\int d^{2}r\psi _{\alpha }^{\dagger }\left( \mathbf{r}\right) K_{\alpha
\beta }\psi _{\beta }\left( \mathbf{r}\right) \mathbf{-}\frac{g}{2}\psi
_{\alpha }^{\dagger }\left( \mathbf{r}\right) \psi _{\beta }^{\dagger
}\left( \mathbf{r}\right) \psi _{\beta }\left( \mathbf{r}\right) \psi
_{\alpha }\left( \mathbf{r}\right) \text{;}  \notag \\
K_{\alpha \beta }\left( \mathbf{\nabla }\right) =\mathcal{E}_{\alpha \beta
}\left( \mathbf{\nabla }\right) -\mu \delta _{\alpha \beta }\text{,}
\label{Hamiltonian}
\end{gather}%
where space is two dimensional, $\mathbf{r=}\left\{ x,y\right\} $ and $\mu $
is the chemical potential. The precise definition of the relevant "single"
electronic excitations $\mathcal{E}_{\alpha \beta }\left( \mathbf{\nabla }%
\right) $ will be dependent on the specific model considered and is
specified below. The index $\alpha $ of the spinors $\psi $ refers to
valley/spin degrees of freedom. The partition function is

\begin{equation}
Z=\mathrm{Tr}\text{ }e^{-H/T}=\int D\psi ^{+}D\psi e^{-S\left[ \psi _{\alpha
}^{+}\mathbf{,}\psi _{\alpha }\right] }\text{,}  \label{Z}
\end{equation}%
with measure defined by independent Grassmann variables $D\psi ^{+}=\underset%
{\alpha }{\Pi }d\psi _{\alpha }^{+},D\psi =\underset{\alpha }{\Pi }d\psi
_{\alpha }$. The Matsubara action reads:%
\begin{eqnarray}
S\left[ \psi ^{+}\mathbf{,}\psi \right] &=&\int_{0}^{1/T}d\tau \int_{r}\psi
_{\alpha }^{+}\left( \tau ,\mathbf{r}\right) \left( \partial _{\tau
}+K_{\alpha \beta }\right) \psi _{\beta }\left( \tau ,\mathbf{r}\right)
\notag \\
&&\mathbf{-}\frac{g}{2}\psi _{\alpha }^{+}\left( \tau ,\mathbf{r}\right)
\psi _{\beta }^{+}\left( \tau ,\mathbf{r}\right) \psi _{\beta }\left( \tau ,%
\mathbf{r}\right) \psi _{\alpha }\left( \tau ,\mathbf{r}\right) \text{,}
\label{S}
\end{eqnarray}%
with the anti-periodic conditions,%
\begin{eqnarray}
\psi _{\alpha }^{+}\left( \tau ,\mathbf{r}\right) &=&-\psi _{\alpha
}^{+}\left( \tau +1/T,\mathbf{r}\right) ,  \label{antiperiod} \\
\psi _{\alpha }\left( \tau ,\mathbf{r}\right) &=&-\psi _{\alpha }\left( \tau
+1/T,\mathbf{r}\right) \text{.}  \notag
\end{eqnarray}%
The local interaction term is not the most general one, but generalization
to more "exotic" local cases (inter-valley\cite{Fu} the exchange spin - spin
coupling\cite{Rosenstein15}) is quite straightforward.

The normal and anomalous Green's functions (consistent with definition in
ref.\cite{AGD}) are:%
\begin{eqnarray}
\left\langle T\psi _{\alpha }\left( X\right) \psi _{\beta }^{+}\left(
X^{\prime }\right) \right\rangle  &=&-G_{\alpha \beta }\left( X\mathbf{;}%
X^{\prime }\right) \text{;}  \notag \\
\left\langle T\psi _{\alpha }\left( X\right) \psi _{\beta }\left( X^{\prime
}\right) \right\rangle  &=&F_{\alpha \beta }\left( X\mathbf{;}X^{\prime
}\right) \text{;}  \label{GF_Gorkov} \\
\left\langle T\psi _{\alpha }^{+}\left( X\right) \psi _{\beta }^{+}\left(
X^{\prime }\right) \right\rangle  &=&F_{\alpha \beta }^{+}\left( X\mathbf{;}%
X^{\prime }\right) \text{,}  \notag
\end{eqnarray}%
where $X=\left( \mathbf{r,}\tau \right) $, $X^{\prime }=\left( \mathbf{r}%
^{\prime },\tau ^{\prime }\right) $. For simplicity, we denote $\left\langle
\psi _{\alpha }\left( X\right) \right\rangle $ as $\psi _{\alpha }\left(
X\right) $, and drop the time ordering operation $T$ in correlators. For
example, $\left\langle \psi _{\alpha }\left( X\right) \psi _{\beta }\left(
X^{\prime }\right) \right\rangle $ stands for $\left\langle T\psi _{\alpha
}\left( X\right) \psi _{\beta }\left( X^{\prime }\right) \right\rangle $.
Their expressions via partition function, Eq.(\ref{S}), are given in
Appendix A.

\subsection{Space and spin rotations symmetries}

Generally a system may be invariant under both the space rotation and the
spin rotation separately. Certain valley symmetries are generally present.
In Weyl semimetals, the action is typically only invariant under the
combined space rotation and spin/valley rotation. The space rotation, $%
\mathbf{r}^{\prime }=\Lambda \mathbf{r}$, acts on a generalized spinor field
as:

\begin{eqnarray}
\psi _{\alpha }^{\prime }\left( \mathbf{r}^{\prime },\tau \right) &=&S\left(
\Lambda \right) _{\alpha \alpha ^{\prime }}\psi _{\alpha ^{\prime }}\left(
\mathbf{r},\tau \right) ,  \label{space_trans} \\
\psi _{\beta }^{\prime +}\left( \mathbf{r}^{\prime },\tau \right) &=&\psi
_{\beta ^{\prime }}^{+}\left( \mathbf{r},\tau \right) S\left( \Lambda
\right) _{\beta ^{\prime }\beta }^{\dagger }\text{.}  \notag
\end{eqnarray}%
The invariance of the action under the transformation, the correlators
satisfy:
\begin{eqnarray}
G_{\alpha \beta }\left( X_{1},X_{2}\right) &=&S\left( \Lambda \right)
_{\alpha \alpha ^{\prime }}G_{\alpha ^{\prime }\beta ^{\prime }}\left(
X_{1},X_{2}\right) S\left( \Lambda \right) _{\beta ^{\prime }\beta
}^{\dagger };  \notag \\
F_{\alpha \beta }\left( X_{1},X_{2}\right) &=&S\left( \Lambda \right)
_{\alpha \alpha ^{\prime }}F_{\alpha ^{\prime }\beta ^{\prime }}\left(
X_{1},X_{2}\right) S\left( \Lambda \right) _{\beta ^{\prime }\beta }^{t}
\label{conditions}
\end{eqnarray}%
or in the matrix form:

\begin{eqnarray}
G\left( X_{1},X_{2}\right) &=&S\left( \Lambda \right) G\left(
X_{1},X_{2}\right) S\left( \Lambda \right) ^{\dagger },  \label{inv_mat} \\
F\left( X_{1},X_{2}\right) &=&S\left( \Lambda \right) F\left(
X_{1},X_{2}\right) S\left( \Lambda \right) ^{t}  \notag
\end{eqnarray}

Assuming that the ground state is homogeneous, $G\left( X,X\right) \equiv
G_{c}$ and $F\left( X,X\right) \equiv g^{-1}\Delta $ are constant matrices
satisfying
\begin{equation}
G_{c}=S\left( \Lambda \right) G_{c}S\left( \Lambda \right) ^{+},\Delta
=S\left( \Lambda \right) \Delta S\left( \Lambda \right) ^{t}\text{.}
\label{localcorr}
\end{equation}%
Let $\Sigma $ be generator of $S\left( \Lambda \right) $. Then
\begin{equation}
\left[ \Sigma ,G_{c}\right] =0,\Sigma \Delta +\Delta \Sigma ^{t}=0
\label{rotation_alg}
\end{equation}%
$G_{c}$ and $\Delta $ also satisfy the following equations:%
\begin{equation}
G_{c}^{+}=G_{c},\Delta ^{t}=-\Delta \text{.}  \label{Pauli}
\end{equation}%
In superconductor $G_{\alpha \beta }^{c}=n_{c}\delta _{\alpha \beta }$ and
leads within the BCS approximation just to renormalization of the chemical
potential. Therefore we finally obtain the Gor'kov equations,
\begin{gather}
\left( -\partial _{\tau }\delta _{\alpha \beta }-K_{\alpha \beta }\left(
\mathbf{\nabla }\right) \right) G_{\beta \gamma }\left( X,X^{\prime }\right)
\mathbf{-}g\times  \notag \\
F_{\beta \alpha }\left( X,X\right) F_{\beta \gamma }^{+}\left( X,X^{\prime
}\right) =\delta \left( X-X^{\prime }\right) \delta _{\alpha \gamma };
\label{GeqX} \\
\left( \partial _{\tau }\delta _{\alpha \beta }-K_{\beta \alpha }\left( -%
\mathbf{\nabla }\right) \right) F_{\beta \gamma }^{+}\left( X,X^{\prime
}\right) -  \notag \\
gF_{\alpha \beta }^{+}\left( X,X\right) G_{\beta \gamma }\left( X,X^{\prime
}\right) =0\text{.}  \notag
\end{gather}

We will also discuss the non-superconducting state like the exciton
condensate with opposite bulk properties, $G_{\alpha \beta }^{c}\left(
X,X\right) \neq n_{c}\delta _{\alpha \beta }$, $F_{\alpha \beta }\left(
X;X^{\prime }\right) =0$. In this case the Dyson - Schwinger form is more
convenient. The gap equation can be recasted (see Appendix A) in matrix form
as

\begin{equation}
G^{-1}=G_{0}^{-1}+g\delta \left( X-X^{\prime }\right) G_{c}^{\prime }\text{,}
\label{DSeq}
\end{equation}%
where $G_{\alpha \beta }^{\prime }$ is the traceless part of $G_{\alpha
\beta }$,
\begin{gather}
G_{\alpha \beta }^{\prime }\left( X;X^{\prime }\right) =-\left\langle \psi
_{\alpha }\left( X\right) \psi _{\beta }^{+}\left( X^{\prime }\right)
\right\rangle _{c}+  \label{traceless} \\
\frac{1}{4}\delta _{\alpha \beta }\sum \left\langle \psi _{\gamma }\left(
X\right) \psi _{\gamma }^{+}\left( X^{\prime }\right) \right\rangle _{c},
\notag
\end{gather}%
with renormalized chemical potential taking case of the trace as in
superconductor.

The Matsubara Green's functions ($\tau $ is the Matsubara time) for uniform
superconducting states can be expressed via Fourier transforms,
\begin{gather}
G_{\alpha \beta }\left( \mathbf{r},\tau ;\mathbf{r}^{\prime },\tau ^{\prime
}\right) =\frac{T}{\left( 2\pi \right) ^{D}}\int d^{D}\mathbf{k}\times
\notag \\
e^{i\mathbf{k\cdot }\left( \mathbf{r-r}^{\prime }\right)
}\sum_{n}e^{-i\omega _{n}\left( \tau -\tau ^{\prime }\right) }G_{\alpha
\beta }\left( \mathbf{k,}\omega _{n}\right) \text{;}  \label{GFdef} \\
F_{\alpha \beta }^{\dagger }\left( \mathbf{r},\tau ;\mathbf{r}^{\prime
},\tau ^{\prime }\right) =\frac{T}{\left( 2\pi \right) ^{D}}\int d^{D}%
\mathbf{k}\times  \notag \\
e^{i\mathbf{k\cdot }\left( \mathbf{r-r}^{\prime }\right)
}\sum_{n}e^{-i\omega _{n}\left( \tau -\tau ^{\prime }\right) }F_{\alpha
\beta }^{\dagger }\left( \mathbf{k,}\omega _{n}\right) \text{,}  \notag
\end{gather}%
where $\omega =\pi T\left( 2n+1\right) $ is the Matsubara Fermionic
frequency. The Matsubara Green's functions in Fourier forms can simplify the
calculation significantly.

General formula\cite{AGD} obtained in Appendix B for the energy of the
superconducting state reads:
\begin{equation}
\frac{d\Omega }{d\left( g^{-1}\right) }\mathcal{=}\frac{V}{2}\text{Tr}\left(
\Delta \Delta ^{\dag }\right)  \label{energy_super}
\end{equation}%
while for chiral symmetry breaking (exciton condensation) states the
gaussian energy is\cite{CJT}
\begin{equation}
\Omega =-\text{Tr}\{-\ln G+[G_{0}^{-1}G-1]\}+\frac{g}{2}\text{Tr}%
G_{c}G_{c}^{\dag }\ \text{.}  \label{energy_chiral}
\end{equation}

\section{The Dirac model and its symmetries}

\subsection{Hamiltonian of the time reversal invariant TI with local
interaction.}

Electrons on the surface of a TI perpendicular to $z$ axis are described by
a Pauli spinors $\psi \left( \mathbf{r}\right) $, where the upper plane, $%
\mathbf{r}=\left\{ x,y\right\} $. In principle there are multiple valleys.
The case of just one valley describing surface of the topological insulator
like $Bi_{2}Te_{3}$ were considered in \cite{Zhang,Herbut,Li14}. It breaks
time reversal invariance and does not allow chiral symmetry breaking, so
here we consider the simplest case of multiple valleys: the Dirac model in
which chiralities of the two Weyl modes are opposite described by field
operators $\psi _{fs}\left( \mathbf{r}\right) $, where $f=L,R$ are the
valley index (pseudospin) for the left/right chirality bands with spin
projections taking the values $s=\uparrow ,\downarrow $ with respect to, for
example, $z$ axis. To use the Dirac ("pseudo-relativistic") notations, these
are combined into a four component bi-spinor creation operator, $\psi
^{\dagger }=\left( \psi _{L\uparrow }^{\dagger },\psi _{L\downarrow
}^{\dagger },\psi _{R\uparrow }^{\dagger },\psi _{R\downarrow }^{\dagger
}\right) $, whose index $\gamma =\left\{ f,s\right\} $ takes four values.
The non-interacting massless Hamiltonian with Fermi velocity $v_{F}$ and
linear dispersion relation, see Fig.1, reads\cite{Wang13},%
\begin{equation}
\mathcal{E}_{\gamma \delta }=-i\hbar v_{F}\nabla ^{i}\alpha _{\gamma \delta
}^{i}\text{,}  \label{epsilon}
\end{equation}%
where two $4\times 4$ matrices, $i=x,y$,
\begin{equation}
\mathbf{\alpha }=\left(
\begin{array}{cc}
\mathbf{\sigma } & 0 \\
0 & -\mathbf{\sigma }%
\end{array}%
\right) \text{,}  \label{alfa_mat}
\end{equation}%
are presented in the block form via Pauli matrices $\mathbf{\sigma }$. They
are related to the Dirac $\mathbf{\gamma }$ matrices (in the chiral
representation, sometimes termed "spinor") by $\ \mathbf{\alpha }=\beta
\mathbf{\gamma }$ with%
\begin{equation}
\beta =\left(
\begin{array}{cc}
0 & \mathbf{1} \\
\mathbf{1} & 0%
\end{array}%
\right) \text{.}  \label{beta_mat}
\end{equation}%
\begin{figure*}[tbp]
\begin{center}
\includegraphics[width=8cm]{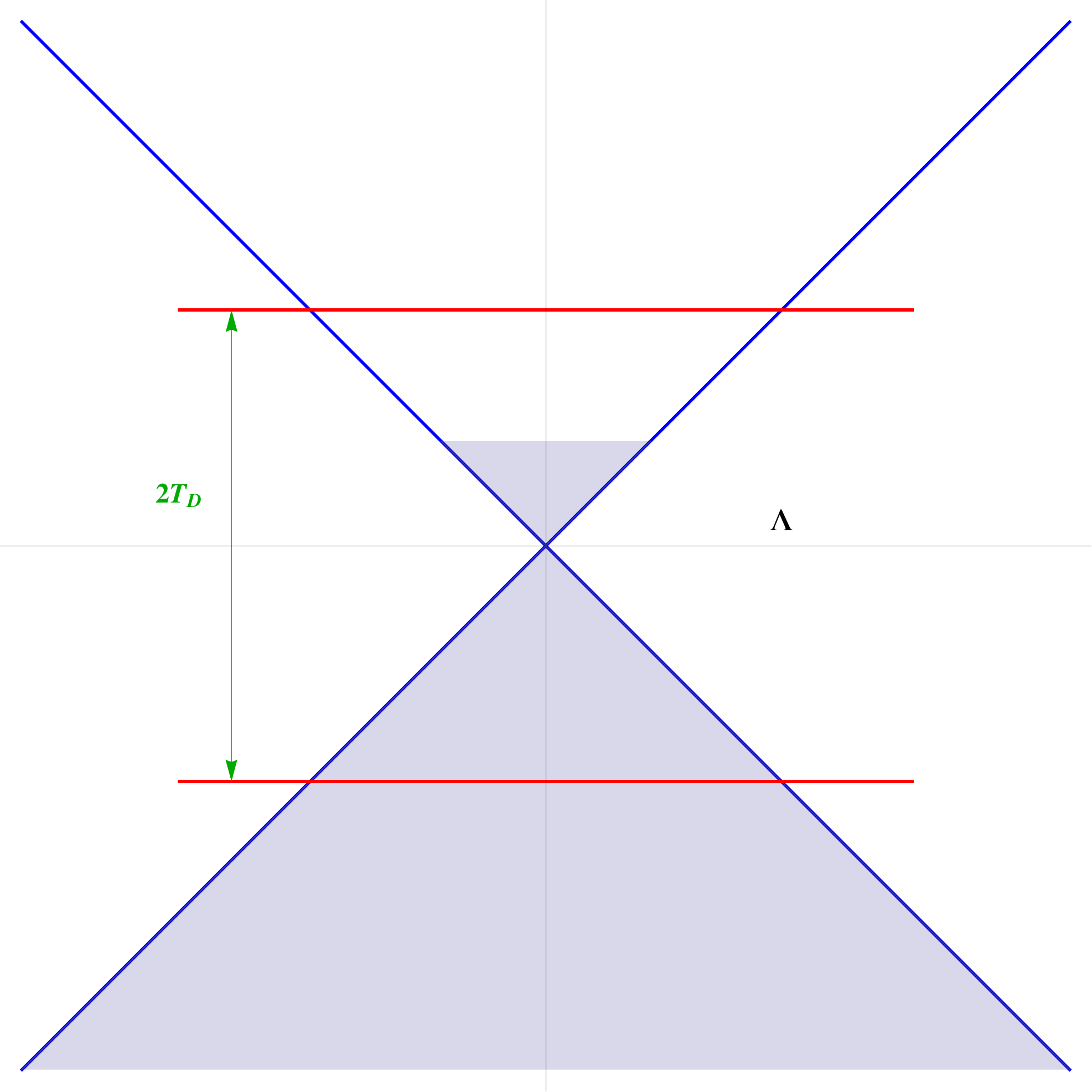}
\end{center}
\par
\vspace{-0.5cm}
\caption{Schematic picture of the band reconstruction of Weyl semi-metal.}
\end{figure*}
The noninteracting Hamiltonian in these notations reads:%
\begin{equation}
K=\int_{\mathbf{r}}\psi ^{\dagger }\left\{ -i\hbar v_{F}\beta \gamma \cdot
\mathbf{\nabla }-\mu \right\} \psi \text{.}  \label{K_Dirac}
\end{equation}%
The Matsubara action, Eq.(\ref{S}), is conveniently written in
pesudo-relativistic notations with $\overline{\psi }=\psi ^{+}\gamma _{0}$
with (Euclidean) $\gamma _{0}=i\beta $:%
\begin{eqnarray}
S &=&\int_{\tau =0}^{1/T}\int_{\mathbf{r}}\left\{ -\overline{\psi }\left(
\gamma _{0}\partial _{\tau }+\hbar v_{F}\gamma \cdot \mathbf{\nabla -}\mu
\gamma _{0}\right) \psi +\right.  \label{relativistic action} \\
&&\left. \frac{g}{2}\overline{\psi }\gamma _{0}\psi \overline{\psi }\gamma
_{0}\psi \right\} \text{.}  \notag
\end{eqnarray}%
Since our focus is on symmetry and its spontaneous breaking, let us review
known discrete and continuous symmetries.

\subsection{Continuous symmetries}

Symmetries of the 2D Dirac model with local interactions Eqs.(\ref%
{Hamiltonian},\ref{epsilon}) were thoroughly discussed in relation to
graphene\cite{Gusynin}. They include parity, time reversal and the discrete
chiral (flavor) transformation: $\psi \rightarrow \gamma _{5}\psi ;\overline{%
\psi }\rightarrow -\overline{\psi }\gamma _{5}$, where $\gamma _{5}=\gamma
_{0}\gamma _{1}\gamma _{2}\gamma _{3}=\gamma _{5}^{\dagger }$. Spontaneous
breaking of this symmetry has been comprehensively investigated in the
context of graphene\cite{Gusynin} and will not be stressed here. There are
also three continuous symmetries that in principle can lead to ordered phase
with massless Goldstone bosons (order parameter waves). The first is the
usual electric charge $U\left( 1\right) $, $\psi \rightarrow e^{i\chi }\psi $%
, that is spontaneously broken in a superconducting state studied in next
section. In addition there is the "chiral" flavour rotations $SU\left(
2\right) $ that play an important role in exciton condensation that will be
addressed in section V.

\subsubsection{Space time symmetries: just Aphelian space rotation combined
with spin (pseudospin) rotation}

Unlike the non-interacting model the pseudorelativistic 2+1 dimensional
Lorentz invariance is explicitly broken by the static interaction Eq.(\ref%
{Hamiltonian}), so that only 2D rotations accompanied by the (pseudo) spin
rotation already described in Section II with spin $\Sigma $ operator,
\begin{equation}
\Sigma \mathbf{=}\left(
\begin{array}{cc}
\sigma _{z} & 0 \\
0 & \sigma _{z}%
\end{array}%
\right) =\gamma _{1}\gamma _{2}\text{,}  \label{spin_Dirac}
\end{equation}%
remain a symmetry. The conserved quantity is the abelian angular momentum
\begin{equation}
J=\int_{\mathbf{r}}\psi ^{\dagger }\left( \mathbf{r}\right) \left\{
i\varepsilon _{ij}r_{i}\nabla _{j}+\frac{1}{2}\mathbf{\Sigma }\right\} \psi
\left( \mathbf{r}\right) \text{.}  \label{J}
\end{equation}%
The second part is referred to as the spin rotation, $S=\frac{1}{2}\int \psi
^{\dagger }\mathbf{\Sigma }\psi $.

\subsubsection{Electric charge $U\left( 1\right) $}

The usual electric charge $U\left( 1\right) $ with conserved electric charge:

\begin{equation}
Q=\int_{\mathbf{r}}\psi ^{\dagger }\left( \mathbf{r}\right) \psi \left(
\mathbf{r}\right) =\int_{\mathbf{r}}\rho \left( \mathbf{r}\right) \text{.}
\label{Q}
\end{equation}%
Action Eq.(\ref{relativistic action}) is invariant under the phase (global
gauge) transformation, $\psi \rightarrow e^{i\chi }\psi $

\subsubsection{Flavour (chiral or valley) $SU(2)$}

It was noticed early on in relation to graphene\cite{Gusynin} that there is
flavour $SU\left( 2\right) $ symmetry. It is shown in Appendix C that
quantities
\begin{equation}
Q_{i}=\int_{\mathbf{r}}\psi ^{\dagger }\left( \mathbf{r}\right) T_{i}\psi
\left( \mathbf{r}\right) ,  \label{Q_i}
\end{equation}%
commute with Hamiltonian and thus are conserved quantities.The generator
matrices,
\begin{equation}
T_{1}=\frac{i}{2}\gamma _{3},T_{2}=\frac{1}{2}\gamma _{5},T_{3}=\frac{1}{2}%
\gamma _{3}\gamma _{5}\text{,}  \label{Tmat}
\end{equation}%
($\gamma _{5}=\gamma _{0}\gamma _{1}\gamma _{2}\gamma _{3}$) constitute a
nonrelativistic $SU\left( 2\right) $ algebra:%
\begin{equation}
\left[ T_{i},T_{j}\right] =i\varepsilon _{ijk}T_{k}\text{.}  \label{SU(2)}
\end{equation}%
A discrete chiral symmetry is just the chiral rotation by angle $\pi $.The
action Eq.(\ref{relativistic action}) is invariant under infinitesimal
transformation, $\delta \psi =iT_{i}\psi ;$ \ $\delta \overline{\psi }=i%
\overline{\psi }\gamma _{0}T_{i}\gamma _{0}$. All three continuous
symmetries commute??(meaning charge, Flavour, and space time symmetry
commute).

\subsection{Spontaneously broken electric charge $U\left( 1\right) $
symmetry phases: superconducting pairing channels}

Due to locality of the dominant interactions, the superconducting order
parameter is local,%
\begin{equation}
O=\int_{\mathbf{r}}\psi _{\alpha }^{\dagger }\left( \mathbf{r}\right)
M_{\alpha \beta }\psi _{\beta }^{\dagger }\left( \mathbf{r}\right) ,
\label{O}
\end{equation}%
where the constant matrix $M$ should be a $4\times 4$ antisymmetric matrix.
Due to the rotation symmetry they transform covariantly under infinitesimal
rotations generated by the spin $\Sigma $ operator Eq.(\ref{spin_Dirac}).

Out of 16 matrices of the four dimensional Clifford algebra six are
antisymmetric. We will not consider rather exotic phases in which in
addition to the charge $U\left( 1\right) $ symmetry neither the 2D rotations
or the $SU\left( 2\right) $ chiral transformations are spontaneously broken.
Therefore superconducting order parameter is invariant under the remaining
symmetries: $\left[ O,J\right] =\left[ O,Q_{i}\right] =0$. Namely it is
invariant under the $SU\left( 2\right) $ and is either scalar or
pseudoscalar under rotations. The requirement of invariance expressed using
Eq.(\ref{spin_Dirac}) takes a form:

\begin{eqnarray}
\left[ O,J\right] &=&\int_{\mathbf{r,r}^{\prime }}\left[ \psi _{\alpha
}^{\dagger }\left( \mathbf{r}\right) M_{\alpha \beta }\psi _{\beta
}^{\dagger }\left( \mathbf{r}\right) \right. ,  \notag \\
&&\left. \psi _{\gamma }^{\dagger }\left( \mathbf{r}^{\prime }\right)
\mathbf{\Sigma }_{\gamma \delta }\psi _{\delta }\left( \mathbf{r}^{\prime
}\right) \right]  \label{transformation} \\
&=&-\int_{\mathbf{r}}\psi ^{\dagger }\left( \mathbf{r}\right) \left( \mathbf{%
\Sigma }M+M\mathbf{\Sigma }^{t}\right) \psi ^{\dagger }\left( \mathbf{r}%
\right) =0\text{.}  \notag
\end{eqnarray}%
Similarly
\begin{equation}
\left[ O,Q_{i}\right] =-\int_{\mathbf{r}}\psi ^{\dagger }\left( \mathbf{r}%
\right) \left( T_{i}M+MT_{i}^{t}\right) \psi ^{\dagger }\left( \mathbf{r}%
\right) =0\text{.}  \label{rotation1}
\end{equation}%
One finds that the only scalar is, $M=i\alpha _{y}$. There is also a
pseudoscalar that will not be discussed here, namely we assume that the
superconducting state preserves all the other symmetries. Which one of the
condensates is realized at zero temperature is determined by the parameters
of the Hamiltonian along the line of dynamical calculation presented in
section IV for the scalar.

It turns out that there when the effective electron attraction is replaced
by repulsion and the superconductivity is not realized there is still a
possibility of continuous symmetry breaking that also belongs to a chiral
universality class: the chiral $SU\left( 2\right) $.

\subsection{Chiral $SU\left( 2\right) $ broken excitonic phases}

In this subsection we consider an opposite situation when the charge
symmetry is unbroken, that is no superconducting condensate appears. Still
due to nontrivial multicomponent situation with the $SU\left( 2\right) $
symmetry there are possible transitions into a gapped exciton condensate
phases. It is plausible that rotational symmetry is also unbroken. Still
there are two possible patterns, one is breaking down to an $U\left(
1\right) $ subgroup with two Goldstone boson modes and another down to
trivial subgroup with three Goldstone models.

General order parameter now is%
\begin{equation}
P=\int_{\mathbf{r}}\psi _{\alpha }^{\dagger }\left( \mathbf{r}\right)
V_{\alpha \beta }\psi _{\beta }\left( \mathbf{r}\right) ,  \label{Pdef}
\end{equation}%
where the constant matrix $V$ should be an $4\times 4$ hermitian matrix.
There are four chiral $SU\left( 2\right) $ triplets of order parameters $%
\mathbf{P=}\left\{ P_{1},P_{2},P_{3}\right\} $ (that can be viewed as the $%
O\left( 3\right) $ vectors). Their commutations with chiral rotations
generators $Q_{i}$ defined in Eq.(\ref{Q_i}) are:
\begin{equation}
\left[ P_{i},Q_{j}\right] =\int_{\mathbf{r}}\psi ^{\dagger }\left( \mathbf{r}%
\right) \left[ V_{i},T_{j}\right] \psi \left( \mathbf{r}\right) =\varepsilon
_{ijk}P_{k}\text{.}  \label{commP}
\end{equation}%
Four sets of matrices $V_{\alpha \beta }$ in terms of the Dirac and chiral
symmetry matrices are%
\begin{eqnarray}
V^{\left( 1\right) } &=&\{\gamma _{0}T_{2},\gamma _{0}T_{1},\frac{1}{2}%
i\gamma _{0}\}=\frac{1}{2}\left\{ -\gamma _{1}\gamma _{2}\gamma _{3},i\gamma
_{0}\gamma _{3},i\gamma _{0}\right\}  \notag \\
V^{\left( 2\right) } &=&\left\{ T_{1},T_{2},T_{3}\right\} =\frac{1}{2}%
\left\{ i\gamma _{3},\gamma _{0}\gamma _{1}\gamma _{2}\gamma _{3},-\gamma
_{0}\gamma _{1}\gamma _{2}\right\}  \notag \\
V^{\left( 3\right) } &=&\left\{ \gamma _{1},\gamma _{1}\gamma _{3},\gamma
_{0}\gamma _{2}\gamma _{3}\right\} ,  \label{Vmat} \\
V^{\left( 4\right) } &=&\left\{ \gamma _{2},\gamma _{2}\gamma _{3},\gamma
_{0}\gamma _{1}\gamma _{3}\right\} \text{.}  \notag
\end{eqnarray}%
There are also four chiral scalars, $\left[ P,Q_{i}\right] =0$, $Q=\int \psi
^{\dagger }I\psi $ (charge)$,S=\int \psi ^{\dagger }\gamma _{1}\gamma
_{2}\psi $ (spin), $H_{i}=\int \psi ^{\dagger }\alpha _{i}\psi ,\alpha
_{i}=\gamma _{0}\gamma _{i}$,$i=x,y$, that complete the Clifford algebra
consisting of 16 hermitian matrices. These are not order parameters and
hence will not be of interest to us. We also limit ourselves to the rotation
invariant phases.\bigskip\ The requirement of invariance expressed using Eq.(%
\ref{spin_Dirac}) takes a form:

\begin{equation}
\left[ P,J\right] =\int_{\mathbf{r}}\psi ^{\dagger }\left( \mathbf{r}\right) %
\left[ V\mathbf{,}\Sigma \right] \psi ^{\dagger }\left( \mathbf{r}\right)
\text{.}  \label{rotation}
\end{equation}%
Since $V^{\left( 3\right) }$ and $V^{\left( 3\right) }$ in Eq.(\ref{Vmat})
are not invariant under rotations only the first two are considered. If only
one of the chiral vector order parameters has a nonzero expectation value,
say $\left\langle P_{3}^{\left( 1\right) }\right\rangle \not=0$, the
symmetry breaking pattern is $SU\left( 2\right) \rightarrow U\left( 1\right)
$, since $\left[ P_{3}^{\left( 1\right) },Q_{3}\right] =0$. According to the
Goldstone theorem there are two soft modes in directions $Q_{1}$ and $Q_{2}$%
. If in addition the second vector order parameter acquires VEV, $%
\left\langle P_{i}^{\left( 2\right) }\right\rangle \not=0$ and $i\not=3$,
the pattern will be $SU\left( 2\right) \rightarrow I$ with three Goldstone
modes.

Which symmetry breaking mode is actually realized at given parameters of the
system (chemical potential, Fermi velocity, interaction sign and strength,
temperature...) is a dynamical question. Therefore now we turn to dynamical
aspects of the phase diagram of the Dirac model.

\section{Superconducting state}

Within gaussian approximation, the Green's functions obey the Gor'kov
equations derived in\cite{Li14} and in last section. For $g>0$ anomalous
Green's functions are nonzero, so that the Gor'kov equations for Fourier
components of the Greens functions simplify considerably,
\begin{eqnarray}
D_{\gamma \beta }^{-1}G_{\beta \kappa }\left( \omega ,p\right) -\widehat{%
\Delta }_{\gamma \beta }F_{\beta \kappa }^{\dagger }\left( \omega ,p\right)
&=&\delta ^{\gamma \kappa }\text{;}  \label{Gorkov_uniform} \\
D_{\beta \gamma }^{-1}F_{\beta \kappa }^{\dagger }\left( \omega ,p\right) +%
\widehat{\Delta }_{\gamma \beta }^{\ast }G_{\beta \kappa }\left( \omega
,p\right) &=&0\text{,}  \notag
\end{eqnarray}%
where $\ D_{\gamma \beta }^{-1}=\left( i\omega -\mu \right) \delta _{\gamma
\beta }-v_{F}\varepsilon _{ij}p_{i}\alpha _{\alpha \beta }^{j}$ where the
chemical potential is renormalized. The matrix gap function can be chosen as
($\Delta $ real)
\begin{equation}
\widehat{\Delta }_{\beta \gamma }=gF_{\gamma \beta }\left( 0\right) =\left(
\begin{array}{cc}
0 & \Delta \\
-\Delta & 0%
\end{array}%
\right) \text{.}  \label{delta}
\end{equation}

These equations are conveniently presented in matrix form (superscript $t$
denotes transposed and $I$ - the identity matrix):
\begin{eqnarray}
D^{-1}G-\widehat{\Delta }F^{\dagger } &=&I\text{;}  \label{matrixeq} \\
D^{t-1}F^{\dagger }+\widehat{\Delta }^{\ast }G &=&0\text{.}  \notag
\end{eqnarray}%
Solving these equations one obtains
\begin{eqnarray}
G^{-1} &=&D^{-1}+\widehat{\Delta }D^{t}\widehat{\Delta }^{\ast }\text{;}
\label{solution} \\
F^{\dagger } &=&-D^{t}\widehat{\Delta }^{\ast }G\text{,}  \notag
\end{eqnarray}%
with the gap function found from the consistency condition
\begin{equation}
\widehat{\Delta }^{\ast }=-g\sum\limits_{\omega q}D^{t}\widehat{\Delta }%
^{\ast }G\text{.}  \label{gap eq}
\end{equation}%
The off-diagonal component of this equation is:
\begin{gather}
\frac{1}{g}=\sum\limits_{\omega p}\left( \Delta ^{2}+v_{F}^{2}p^{2}+\mu
^{2}+\hbar ^{2}\omega ^{2}\right) \times  \notag \\
\left( \Delta ^{2}+\hbar ^{2}\omega ^{2}+\left( v_{F}p-\mu \right)
^{2}\right) ^{-1}\times  \label{gap} \\
\left( \Delta ^{2}+\hbar ^{2}\omega ^{2}+\left( v_{F}p+\mu \right)
^{2}\right) ^{-1}\text{.}  \notag
\end{gather}

The spectrum of elementary excitations obtained from the poles of the Greens
function coincides with that found within the Bogoliubov - de Gennes
approach \cite{Herbut}: $E_{p}=\pm \sqrt{\Delta ^{2}+\left( v_{F}p-\mu
\right) ^{2}}$.

\subsection{Zero temperature phase diagram for the superconductor - normal
transition.}

At zero temperature the integrations over frequency and momentum limited by
the UV cutoff $\Lambda $ result in
\begin{equation}
U=\sqrt{\Delta ^{2}+\mu ^{2}}-\frac{\mu }{2}\log \frac{\sqrt{\Delta ^{2}+\mu
^{2}}+\mu }{\sqrt{\Delta ^{2}+\mu ^{2}}-\mu }\text{,}  \label{gapeq1}
\end{equation}%
where the dependence on the cutoff is incorporated in the renormalized
coupling with dimension of energy defined as
\begin{equation}
U=v_{F}\Lambda -\frac{4\pi \hbar ^{2}v_{F}^{2}}{g}\text{.}  \label{gren}
\end{equation}%
This can be interpreted as an effective binding energy of the Cooper pair in
the Weyl semi - metal. We consider only $\mu >0$, since the particle - hole
symmetry makes the opposite case of the hole doping, $\mu <0$, identical. Of
course the superconducting solution exists only for $g>0$. In Fig. 2 the
dependence of the gap $\Delta $ as function of the chemical potential $\mu $
is presented for different values of $U$.

\begin{figure}[tbp]
\centering
\includegraphics[width=8cm]{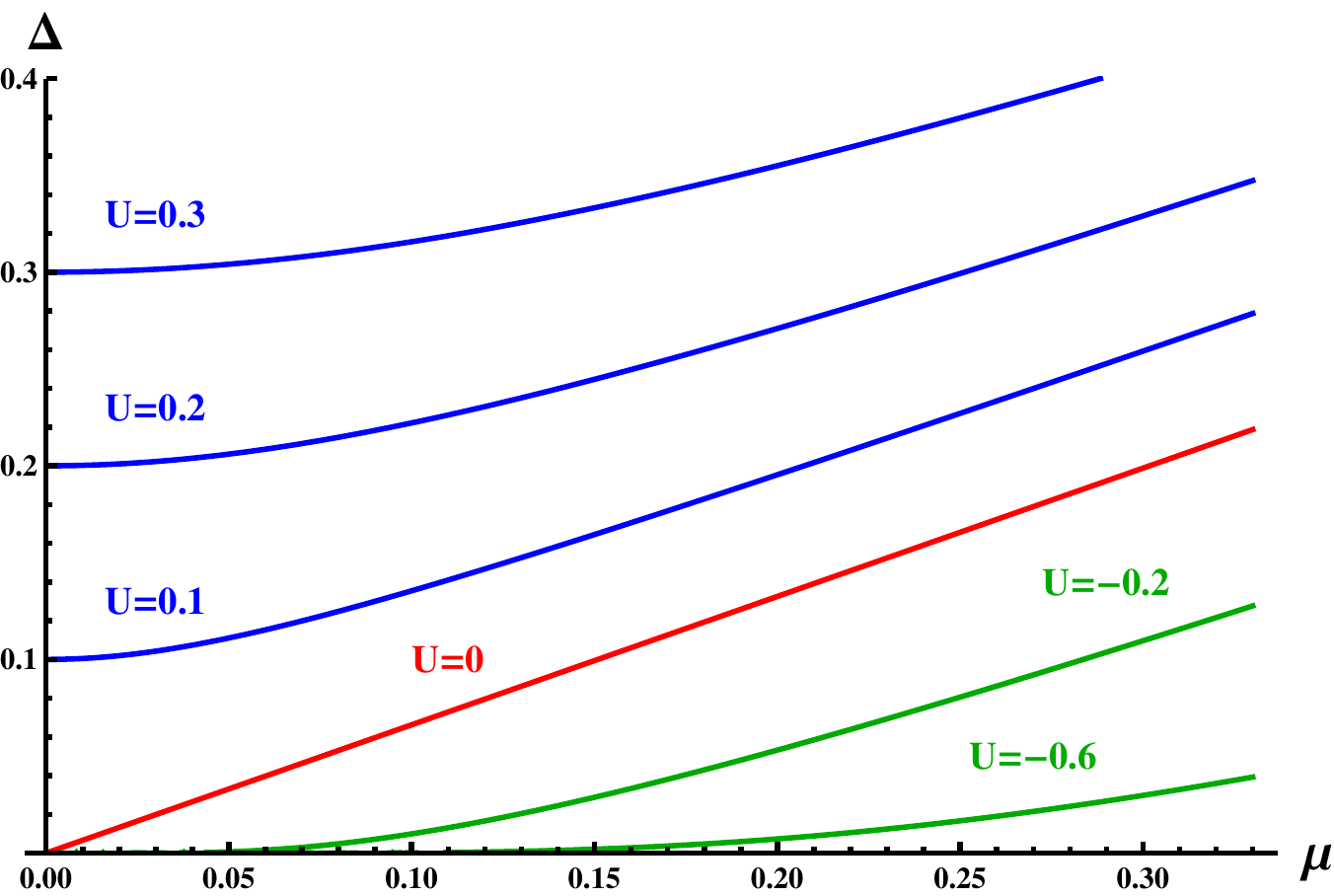}
\caption{Order parameter at zero temperature as function of chemical
potential of the TI surface Weyl semi-metal at various values of coupling
parametrized by the renormalized energy $U$, Eq.(\protect\ref{gren}). For
positive $U$ (blue lines) the superconductivity is strong and does not
vanish even for zero chemical potential. There exists the critical coupling,
$U=0$ (red line), at which the second order transition occurs at quantum
critical point $\protect\mu =0$. For negative $U$ the superconductivity
still exists at $\protect\mu >0$, but is exponentially weak. }
\end{figure}

For an attractive coupling $g$ stronger than the critical one,
\begin{equation}
g_{c}=\frac{4\pi \hbar ^{2}v_{F}}{\Lambda }\text{,}  \label{g_c}
\end{equation}%
(when $U>0$), blue lines in Fig. 2, there are two qualitatively different
cases.

(i). When $\mu <<U$ the dependence of $\Delta $ on the chemical potential is
parabolic, see \cite{Li14}. In particular, when $\mu =0,$ the gap equals $U$%
. As can be seen from Fig. 2, the chemical potential makes a very limited
impact in the large portion of the phase diagram.

(ii) For the attraction just stronger than critical, $g>g_{c}$, namely for
small positive $U$, the dependence becomes linear, see red line in Fig. 2, $%
\Delta =0.663\,\mu $. So that the already weak condensate becomes sensitive
to $\mu $.

The case (i) is more interesting than (ii) since it exhibits stronger
superconductivity (larger $T_{c}$, see below). Finally for $g<g_{c}\,$,
namely negative $U$ (green lines), the superconductivity is very weak with
exponential dependence similar to the BCS one, $\Delta \approx \mu $ exp$%
\left[ -\left( \left\vert U\right\vert /\mu -1\right) \right] $. As was
mentioned above, in the more interesting cases of large $\Delta $ the
dependence on the chemical potential is very weak. A peculiarity of
superconductivity in TI is that electrons (and holes) in Cooper pairs are
created themselves by the pairing interaction rather than being present in
the sample as free electrons. Therefore it is shown that it is possible to
neglect the effect of weak doping and consider directly the $\mu =0$
particle-hole symmetric case. This point in parameter space is the QCP \cite%
{Sachdev} and will be studied in detail in what follows. Of course, at
finite temperature at any attraction, $g>0$, there exists a (classical)
superconducting critical point at certain temperature $T_{c}$ that is
calculated next.

\subsection{Dependence of the critical temperature $T_{c}$ on strength of
pairing interaction.}

Summation over Matsubara frequency and integrations over momenta in the gap
equation, Eq.(\ref{gapeq1}), at finite temperature and arbitrary chemical
potential. The critical temperature as a function of $\mu $ and (positive) $%
U $ is obtained numerically and presented in Fig. 3. Again at relatively
large $U$ the dependence of $T_{c}$ on the chemical potential is very weak
and parabolic. When $0<g<g_{c}$ the critical temperature is exponentially
small albeit nonzero, $\frac{\Delta }{U}\approx 1+\left( \frac{\mu }{U}%
\right) ^{2} $.
\begin{figure}[tbp]
\centering
\includegraphics[width=8cm]{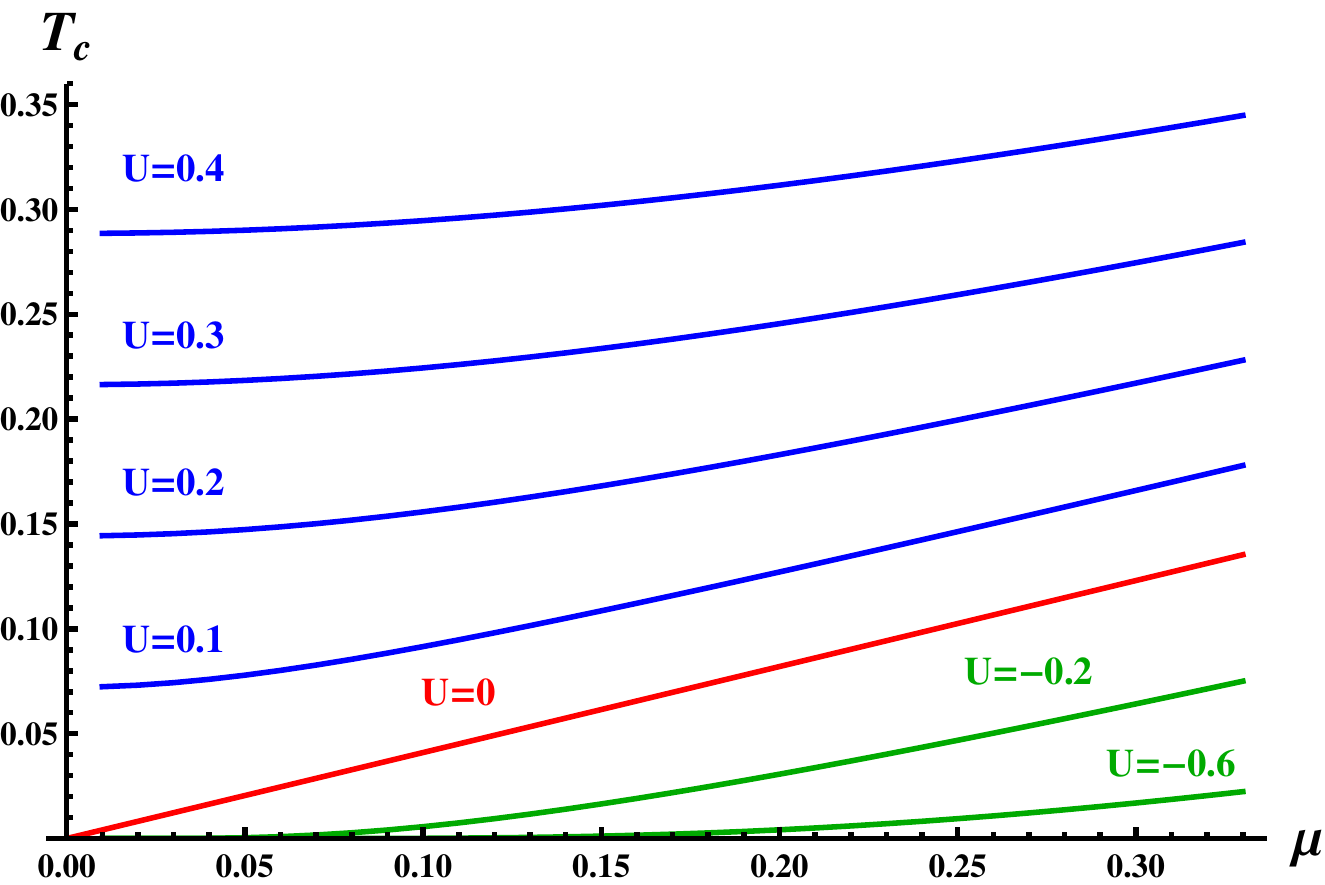}
\caption{Transition temperature as function of chemical potential at
supercritical ($U>0$, in blue), critical ($U=0$, in red) and subcritical
values of coupling. }
\end{figure}

\subsection{Zero chemical potential $\protect\mu =0$.}

At zero chemical potential the Hamiltonian Eq.(\ref{Hamiltonian}) possesses
a particle - hole symmetry. Microscopically, Cooper pairs of both electrons
and holes are formed, see Fig. 1a. The system is unique in this sense since
the electron - hole symmetry is not spontaneously broken in both normal and
superconducting phases. Supercurrent in such a system does not carry
momentum or mass. Performing the sum and integral over momenta in the gap
equation, Eq.(\ref{gapeq1}), analytically (see Appendix A), it becomes
(using the definition of $U$ given in Eq.(\ref{gren})) for $U>0$:
\begin{equation}
U=2T\log \left[ 2\cosh \frac{\Delta }{2T}\right] \text{.}  \label{gapren_T}
\end{equation}%
At zero temperature $\Delta =U$, while $\Delta \rightarrow 0$ as a power of
the parameter $U\propto g-g_{c}$ describing the deviation from quantum
criticality
\begin{equation}
T_{c}=\frac{1}{2\log 2}U^{z\nu };\text{ \ }z\nu =1\text{.}  \label{Tc_mic}
\end{equation}%
Here $z$ is the dynamical critical exponent\cite{Sachdev}. Therefore, as
expected, the renormalized coupling describing the deviation from the QCP is
proportional to the temperature at which the created condensate disappears.

\begin{figure}[tbp]
\centering
\includegraphics[width=8cm]{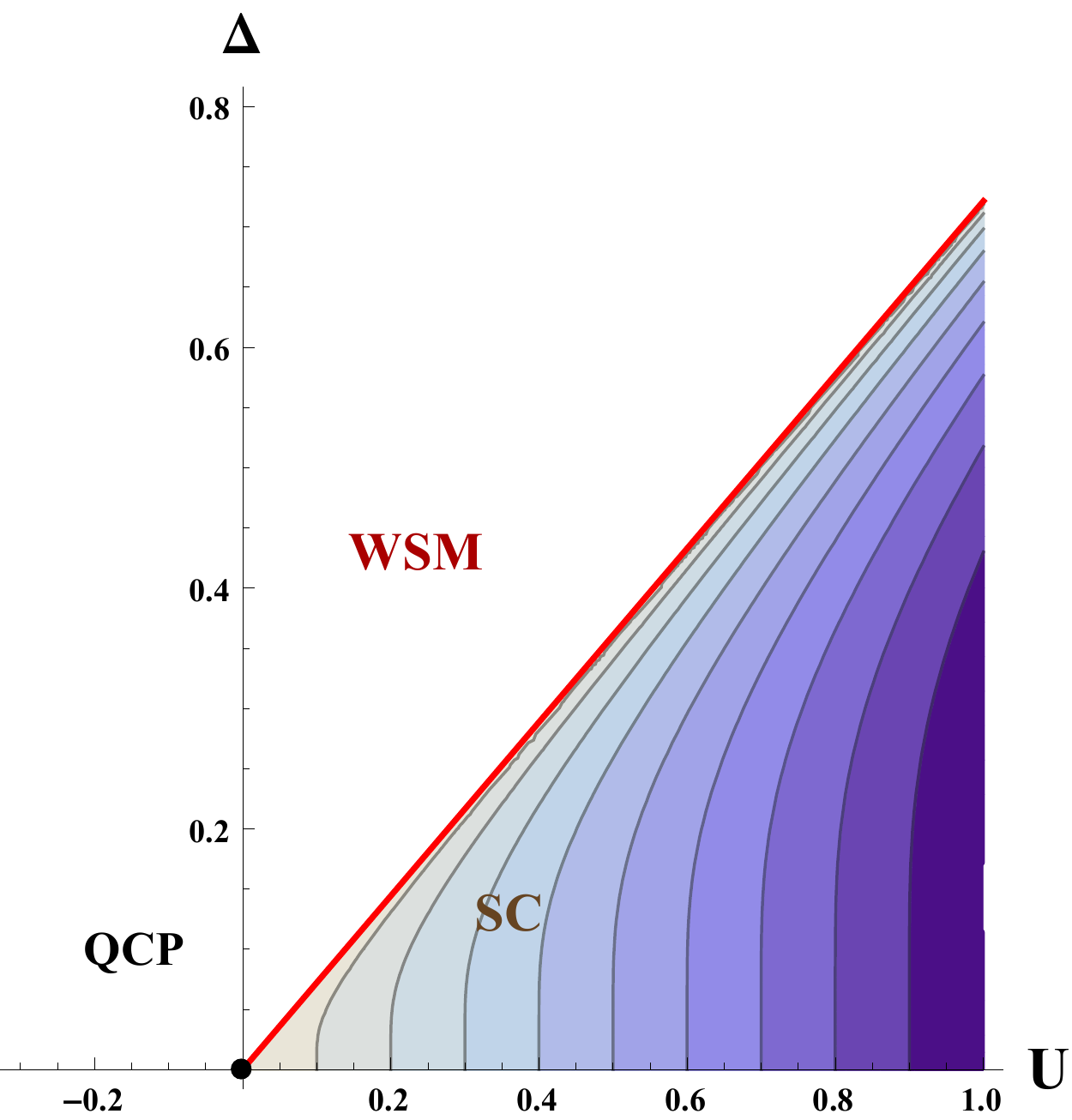}
\caption{Phase diagram of \ STI. Order parameter as function of $U$
describing the deviation from criticality near the quantum critical point at
$\Delta=0$, $\protect\mu =0$. The critical line is a strait line in mean
field approximation.}
\end{figure}

The temperature dependence of the gap reads:
\begin{equation}
\Delta \left( T\right) =2T\cosh ^{-1}\left( \frac{1}{2}\exp \frac{U}{2T}%
\right) \text{.}  \label{delta_mic}
\end{equation}%
This it typical for chiral universality classes \cite{Sachdev,Rosenstein}.

It is interesting to compare this dependence with the conventional BCS\cite%
{AGD} for transition at finite temperature, namely away from QCP, see Fig
1b. At zero temperature $\Delta \left( 0\right) /T_{c}=2\log 2$ $%
=\allowbreak 1.\,\allowbreak 39$ (within BCS - $1.76$), while near $T_{c}$
one gets $\Delta /T_{c}=2^{3/2}\log ^{1/2}2\sqrt{1-t}=$ $2.\,\allowbreak 35%
\sqrt{1-t}$ (BCS - $3.07\sqrt{1-t}$), where $t=T/T_{c}$. To describe the
behavior of the STI in inhomogeneous situations like the external magnetic
field, boundaries, impurities or junction with metals or other
superconductors, it is necessary to derive the effective theory in terms of
the order parameter $\Delta \left( \mathbf{r}\right) $, where $\mathbf{r}$
varies on the mesoscopic scale.

Phase diagram of \ STI order parameter as function of $U$ at zero
temperature is plotted in Fig.4.

\subsection{Coherence length and the condensation energy}

The quadratic term of the Ginzburg-Landau energy $F_{2}=\sum_{\mathbf{p}%
}\Delta _{p}^{\ast }\Gamma \left( p\right) \Delta _{p}$ is obtained exactly
from expanding the gap equation to linear terms in $\Delta $ for arbitrary
external momentum. The dependence on $\mathbf{p}$ is non-analytic and within
our approximation higher powers of $p$ do not appear. The second term is
very different from the quadratic term in the GL functional for conventional
phase transitions at finite temperature \cite{Herbutbook} or even quantum
phase transitions in models without Weyl fermions \cite{Sachdev} and has a
number of qualitative consequences. Comparing the two terms in Eq.(\ref%
{Gamma1}), one obtains the coherence length as a power of parameter $%
U\propto g-g_{c}$ describing the deviation from criticality:
\begin{equation}
\xi \left( U\right) =\frac{\pi }{4}v_{F}\hbar U^{-\nu }\text{; \ \ \ \ }\nu
=1\text{.}  \label{coherence}
\end{equation}%
This is different from the dependence in non-chiral universality classes
that is \cite{Herbutbook} $\xi \left( T\right) \infty \left( T_{c}-T\right)
^{-\nu },$ $\nu =1/2$ in mean field. Of course in the regime of critical
fluctuations this exponent is corrected in both non-chiral \cite{Herbutbook}
and chiral\cite{Gat} universality classes.

Local terms in the GL energy density are also calculable exactly. Expression
for the kernel can be written as a trace:
\begin{gather}
\Gamma =\frac{1}{2}tr\left\{ \sum\limits_{\omega q}\sigma ^{y}D_{\omega
q}^{t}\sigma ^{y}D_{\omega ,q-p}+\frac{1}{g}I\right\}  \label{C1} \\
=\frac{1}{g}-\sum\limits_{\omega q}\frac{\hbar ^{2}\omega
^{2}-v_{F}^{2}p\cdot q+v_{F}^{2}q^{2}}{\left( \hbar ^{2}\omega
^{2}+v_{F}^{2}q^{2}\right) \left( \hbar ^{2}\omega ^{2}+v_{F}^{2}\left\vert
\mathbf{q-p}\right\vert ^{2}\right) }.  \notag
\end{gather}%
Integrating over $\omega $ (at zero temperature) one obtains%
\begin{eqnarray}
\Gamma &=&-\frac{1}{8\pi ^{2}\hbar ^{2}v_{F}}\int_{q,\phi }\frac{pq\cos \phi
-p^{2}}{\left\vert \mathbf{q-p}\right\vert ^{2}+q\left\vert \mathbf{q-p}%
\right\vert }  \label{Gamma} \\
&&-\frac{U}{4\pi \hbar ^{2}v_{F}^{2}},  \notag
\end{eqnarray}%
where $\phi $ is an angle between $\mathbf{q}$ and $\mathbf{p}$. The
integral is homogeneous in momentum and therefore is linear in $p=\left\vert
\mathbf{p}\right\vert $ and one arrives at:
\begin{equation}
\Gamma \left( p\right) =-\frac{U}{4\pi \hbar ^{2}v_{F}^{2}}+\frac{\left\vert
p\right\vert }{16v_{F}\hbar ^{2}}\text{.}  \label{Gamma1}
\end{equation}

\subsection{\ Local terms in Ginzburg - Landau equation and energy}

For $p=0$ the gap equation Eq.(\ref{gapeq1}) reads%
\begin{equation}
\frac{\Delta }{4\pi \hbar ^{2}v_{F}^{2}}\left( -U+\sqrt{\Delta ^{\ast
}\Delta }\right) =0\text{,}  \label{C3}
\end{equation}%
This is obtained from the energy functional%
\begin{equation}
F=\frac{1}{4\pi \hbar ^{2}v_{F}^{2}}\int d^{2}\mathbf{r}\left\{ -U\Delta
^{\ast }\Delta +\frac{2}{3}\left( \Delta ^{\ast }\Delta \right)
^{3/2}\right\} \text{.}  \label{C4}
\end{equation}%
It is quite nonstandard compared to customary quartic term $\left( \Delta
^{\ast }\Delta \right) ^{2}$ in conventional universality classes. The GL
equations in the homogeneous case for the condensate gives $\Delta
_{0}=U^{\beta }$ with critical exponent $\beta =1,$ different from the mean
field value $\beta =1/2$ for the $U\left( 1\right) $ universality class\cite%
{Herbutbook}. The condensation energy density is $f_{0}=-\frac{1}{12\pi
\hbar ^{2}v_{F}^{2}}U^{2-\alpha }$ with $\alpha =-1$. The free energy
critical exponent at QCP therefore is also different from the classical $%
\alpha =0$. The magnetic field couples to the order parameter field by a
standard minimal substitution. The Ginzburg-Landau approach the only
available practical tool to study properties of inhomogeneous configurations%
\cite{Abrikosov} in external magnetic field like the Abrikosov vortex
systems.

\section{Exciton condensation}

\subsection{Chiral symmetry breaking}

Assuming that the electric charge $U\left( 1\right) $ and rotation symmetry
is unbroken (no superconductivity) for insufficiently strong attractive
interaction (pairing) one still can have transitions due to always existing
effective repulsion. This possibility was already considered in graphene\cite%
{Gusynin}. Here we note that the symmetry patten might be very different
from often invoked relativistic $2+1$ Gross - Neveu model thoroughly studied
as a toy model in relativistic quantum field theory. Possible chiral
symmetry breaking states were reviewed in subsection IId and we consider
first a ground state with nonzero order parameter
\begin{equation}
\left\langle P_{3}^{\left( 1\right) }\right\rangle =v\not=0\text{.}
\label{vdef}
\end{equation}%
It is convenient to parametrize the gaussian variational ground state by the
trace of propagator:

\begin{equation}
G\left( X;X\right) =m/g\text{ }\gamma _{0}\text{,}  \label{mdef}
\end{equation}%
with "mass" $m$ determining the order parameter. According to gap equation
derived in Appendix B, we have
\begin{eqnarray}
G^{-1} &=&\left( i\omega +\mu \right) I-\alpha \cdot \left( \hbar
v_{F}k\right) -m\gamma _{0};  \label{DSE} \\
G &=&-\frac{\left( \mu +i\omega \right) I+\alpha \cdot \left( \hbar
v_{F}k\right) +m\gamma _{0}}{m^{2}+\left( \hbar v_{F}k\right) ^{2}-\left(
\mu +i\ \omega \right) ^{2}\ }\text{,}  \notag
\end{eqnarray}%
leads to
\begin{gather}
mI=gG^{\prime }\left( \tau ,\mathbf{r};\tau ,\mathbf{r}\right) \gamma
_{0}^{-1}=-\frac{g}{\left( 2\pi \right) ^{3}}\times  \label{gapchiral} \\
\int_{\omega ,\mathbf{k}}\frac{m}{m^{2}+\left( \hbar v_{F}\mathbf{k}\right)
^{2}-\left( \mu +i\ \omega \right) ^{2}\ }\text{.}  \notag
\end{gather}%
Performing integrations with momentum cutoff $\Lambda $ one obtains for $%
m\not=0$:%
\begin{equation}
1=-\frac{g}{4\pi \left( \hbar v_{F}\right) ^{2}}\left\{
\begin{array}{c}
\left( \sqrt{m^{2}+\Lambda ^{2}}-m\right) ,\mu \leq m \\
\left( \sqrt{m^{2}+\Lambda ^{2}}-\mu \right) ,\mu >m%
\end{array}%
\right. \text{.}  \label{simplified gapeq}
\end{equation}%
For $\mu =m$ one obtains the first order transition point (for $\mu >0$)

\begin{equation}
\text{ }\frac{4\pi \left( \hbar v_{F}\right) ^{2}}{\left\vert
g_{c}\right\vert }=\sqrt{\mu ^{2}+\Lambda ^{2}}-\mu \text{,}
\label{Tc_chiral}
\end{equation}%
For larger repulsion, $\left\vert g\right\vert \geqslant \left\vert
g_{c}\right\vert $, the mass is larger:%
\begin{equation}
m=\mu +\frac{1}{\left( \hbar v_{F}\right) ^{2}}\left( \left\vert
g\right\vert -\left\vert g_{c}\right\vert \right) \left( \frac{\Lambda ^{2}}{%
8\pi }+\frac{2\pi \left( \hbar v_{F}\right) ^{4}}{\left\vert g\right\vert
\left\vert g_{c}\right\vert }\right) .  \label{largerm}
\end{equation}

The free energy, using the Abrikosov formula derived in Appendix B, is ($%
\Lambda >>m$)
\begin{gather}
\left( \Omega \left( m\right) -\Omega _{0}\left( m=0\right) \right)
/V=\int_{0}^{m}2m^{2}\frac{d\frac{1}{\left\vert g\right\vert }}{dm}dm
\label{energy_chiral1} \\
=-\int_{0}^{m}\frac{m^{2}}{2\pi \left( \hbar v_{F}\right) ^{2}}dm=-\frac{%
m^{3}}{6\pi \left( \hbar v_{F}\right) ^{2}}\text{ .}  \notag
\end{gather}%
For $\mu >m$, since $\frac{d\left( 1/\left\vert g\right\vert \right) }{dm}=%
\frac{1}{4\pi \left( \hbar v_{F}\right) ^{2}}\frac{m}{\sqrt{m^{2}+\Lambda
^{2}}}>0$, the chiral symmetry breaking state will have higher energy than
the normal one. For $\left\vert g\right\vert <\left\vert g_{c}\right\vert ,$
the stable ground state is the normal state with $m=0$.

Finite temperature properties including the phase diagram and the QCP at
zero chemical potential can be studied along the lines similar to the
superconducting transition.

\section{Discussion and conclusions}

Having considered both the superconducting and the excitonic transitions for
sufficiently strong attraction or repulsion within the gaussian
approximation, a natural question is what happens when we approach the
quantum critical point. At criticality various renormalization group methods
should be used\cite{Herbutbook}.

\subsection{Criticality beyond the gaussian approximation}

Critical (quantum) fluctuations are expected to be significant in this
relatively low dimensional (relativistic 2+1 dimensional system) system.
Generally they are not as strong as in 2D statistical system at finite
temperature, but stronger than in 3D one. The approximation we have made
describes reasonably well "gaussian" fluctuation beyond the region where
stronger critical fluctuations in these systems appear and should be treated
nonperturbatively\cite{Rosenstein} typically using variants of the
renormalization group approach\cite{Herbutbook}. The critical exponents in
this region differ from the one called "quantum gaussian (BCS)" in ref. \cite%
{Sachdev} and available results are obtained using either $\varepsilon $
expansion\cite{Gat,Herbut09} ($\varepsilon =4-d$, where $d=2+1$ is the
space-time dimension), $1/N$, where $N$ is the number of fermionic species
on the surface\cite{Gat} and functional (strong coupling) RG\cite{Janssen}
and Monte Carlo simulations\cite{old,MC} (with reservations specified
below). The universality class of the supconducting transition according to
classification proposed in ref. \cite{Gat} is the chiral XY (symmetry of
order parameter $U\left( 1\right) $) with $N=1$. The large $N$ expansion is
not reliable for the one component system considered here (but the number
might be larger in similar systems for which our approach trivially
generalizes), so let us use the $\varepsilon $ expansion.

Using the formulas for the anomalous dimensions of the order parameter (see
second reference in \cite{Gat}),
\begin{equation}
\gamma _{\Delta }=\eta =1/4\varepsilon +0.044\varepsilon ^{2}+O\left(
\varepsilon ^{3}\right) \approx 0.294  \label{gamma}
\end{equation}%
and its square,
\begin{equation}
\gamma _{\Delta ^{2}}=\left( 1+\sqrt{11}\right) /10\varepsilon
+0.065\varepsilon ^{2}\simeq 0.43\varepsilon +0.065\varepsilon ^{2},
\label{gamma2}
\end{equation}%
critical exponents are obtained from the hyperscaling relations:
\begin{eqnarray}
\alpha &=&2-d/\left( 2-\gamma _{\Delta ^{2}}\right) =\left( \varepsilon
-2\gamma _{\Delta ^{2}}\right) /\left( 2-\gamma _{\Delta ^{2}}\right)
=-0.353,  \notag \\
\beta &=&\left( 1+\gamma _{\Delta }\right) /2\left( 2-\gamma _{\Delta
^{2}}\right) =0.515\text{.}  \label{alfabeta}
\end{eqnarray}%
and can be compared with those in Table 1 in ref.\cite{Li14}.$\ $The
exponents from the $\varepsilon $ expansion were found to be consistent for
larger values of $N$ with the latest Monte Carlo simulations \cite{MC},
while consistent with the functional RG\cite{Janssen}. The critical
exponents of the chiral transition belongs to this class with $N=2$.

The corresponding chiral universality class for the exciton condensation in
Dirac semi-metal is the Heisenberg $N=2$ ($SU\left( 2\right) $). The
critical exponents for this case were also calculated in ref.\cite%
{Gat,Herbut09,Janssen} Recently they were invoked in a discussion of second
order quantum transitions in Hubard model on honeycomb lattice\cite{Sorella}%
. It should be noted that the flavour symmetries are often broken
"explicitly" by some kind of anisotropy. In this case Goldstone bosons
acquire a small mass (like pions in quantum chromodynamics in which the
chiral symmetry is slightly broken by the light quark masses), although
their major properties remain intact. This can be taken into account as a
small perturbation\cite{Chiraldyn}.

\subsection{Experimental feasibility of observation of quantum phase
transition}

The best candidate to observe the superconductivity is a topological
insulator of the $Bi_{2}Se_{3}$ family. To estimate the pairing efficiency
due to phonons, one should rely on recent studies of surface phonons in TI
\cite{DasSarma}. The coupling constant in the Hamiltonian, Eq.(\ref%
{Hamiltonian}), is obtained from the exchange of acoustic (Rayleigh) surface
phonons $g=\lambda v_{F}^{2}\hbar ^{2}/2\pi \mu $, where $\lambda $ is the
dimensionless effective electron - electron interaction constant of order $%
0.1$ (somewhat lower values are obtained in ref.\cite{Guinea}). It was shown
in ref. \cite{DasSarma} that at zero temperature the ratio of $\lambda $ and
$\mu $ is constant with well defined $\mu \rightarrow 0$ limit with value $%
g=0.23$ $eV\ nm^{2}$ for $v_{F}\approx 7\ \cdot 10^{5}m/s$ (for $%
Bi_{2}Se_{3} $). The critical coupling constant $g_{c}$, Eq.(\ref{g_c}), can
be estimated from the Debye cutoff $T_{D}=200K$ determining the momentum
cutoff $\Lambda =T_{D}/c_{s}$, where $c_{s}$ is the sound velocity. Taking
value to be $c_{s}=2\cdot 10^{3}m/s$ (for $Bi_{2}Se_{3}$), one obtains $%
g_{c}=4\pi v_{F}c_{s}\hbar ^{2}/\,T_{D}=0.20$ $eV$ $nm^{2}$.

Of course the Coulomb repulsion might weaken or even overpower the effect of
the attraction due to phonons, so that superconductivity does not occur. In
TI like $Bi_{2}Se_{3}$ however, the dielectric constant is very large $%
\varepsilon =50$, so that the Coulomb repulsion is weak. Moreover it was
found in graphene (that has identical Coulomb interaction), that although
the semi-metal does not screen \cite{CastroNeto}, the effects of the Coulomb
coupling are surprisingly small, even in leading order in perturbation
theory. The superconductivity was observed in these systems that howeved had
to be either doped in the bulk or on the surface\cite{Koren} (by a $Cu$) or
by applying pressure\cite{pressureBiSe}. It is not yet clear whether the
observed superconductivity is a bulk or a surface effect.

\textit{\ }The Dirac semimetal in optically trapped cold atoms\cite{cold}
offers a well controllable system in which this phenomenon occurs both for
repulsive interaction (chiral symmetry breaking) and in particular the
attractive one (superconductivity) becuase there is no Coulomb repulsion as
the atoms are neutral.

Recently after experimental discovery of 3D Dirac semi-metals\cite{Potemski}
the new class of questions similar to those discussed in present paper
arise. Extraordinary electronic properties of these Dirac materials\cite%
{3Dtheory} including superconductivity\cite{Cava} and chiral condensate are
being studied theoretically and experimentally.

\subsection{Conclusions}

We have studied continuous phase transitions in a Dirac semi-metal realized
recently as a surface of topological insulator. The noninteracting system is
characterized by (nearly) zero density of states on the 2D Fermi manifold.
It degenerates into a point when the chemical potential coincides with the
Weyl point of the surface states as in the original proposal for a major
class of such materials\cite{Zhang1}. The pairing attraction (the most
plausible candidate being surface phonons) therefore has two tasks in order
to create the superconducting condensate. The first is to create a pair of
electrons (that in the present circumstances means creating two holes as
well) and the second is to pair them. To create the charges does not cost
much energy since the spectrum of the Weyl semimetal is gapless (massless
relativistic fermions); this is effective as long as the coupling $g$ is
larger than the critical $g_{c}$, see Eq.(\ref{g_c}). The situation is more
reminiscent of the creation of the chiral condensate in relativistic
massless four - fermion theory (a 2D version\cite{Rosenstein} was recently
contemplated for graphene \cite{CastroNeto,Katsnelson}) than to the BCS or
even BEC in condensed matter systems with parabolic dispersion law. Due to
the special "ultra-relativistic" nature of the pairing transition at zero
temperature as a function of parameters like the pairing interaction
strength is unusual: even the mean field critical exponents are different
from the standard ones that generally belong to the $U\left( 1\right) $
class of second order phase transitions.

To summarize, we studied the phase diagram of the superconducting and chiral
transition at arbitrary chemical potential, effective local interaction
strength and temperature $T$. The quantum ($T=0$) critical point appears at
zero chemical potential and belongs the $U_{N}\left( 1\right) $ chiral
universality class (the subscript denotes number of massless fermions at QCP
according to classification in \cite{Gat,Sachdev}) for the attraction
(superconductivity) and $SU_{N}\left( 2\right) $ for repulsion (exciton
condensation).

\textit{Acknowledgements.} We are indebted to C.W. Luo, J.J. Lin and W.B.
Jian for explaining details of experiments, and T. Maniv and M. Lewkowicz
for valuable discussions. Work of D.L. and B.R. was supported by NSC of
R.O.C. Grants No. 98-2112-M-009-014-MY3 and MOE ATU program. The work of
D.L. also is supported by National Natural Science Foundation of China (No.
11274018). B.R. is grateful to School of Physics of Peking University for
hospitslity.

\appendix

\section{Path integral derivation of the Gorkov and Dyson -
Schwinger equations}

\subsection{General correlations and sources}

\ We introduce the grassmanian source terms into Matsubara action Eq.(\ref{S}%
):

\begin{gather}
S\left[ \psi ^{+}\mathbf{,}\psi ,J^{+},J\right] =S\left[ \psi ^{+}\mathbf{,}%
\psi \right]  \label{A1} \\
-\int_{0}^{\beta }d\tau \int_{r}\left[ \psi _{\sigma }^{+}\left( \tau ,%
\mathbf{r}\right) J_{\sigma }\left( \tau ,\mathbf{r}\right) +\psi _{\sigma
}\left( \tau ,\mathbf{r}\right) J_{\sigma }^{+}\left( \tau ,\mathbf{r}%
\right) \right] \text{.}  \notag
\end{gather}%
The generating functional\ for the disconnected correlations is $Z\left(
J^{+},J\right) $

\begin{eqnarray}
\left\langle \psi _{\alpha }\left( X\right) \right\rangle  &=&\frac{1}{Z}%
\frac{\delta Z}{\left[ -\delta J_{\alpha }^{+}\left( X\right) \right] }\text{%
;}  \label{A2} \\
\left\langle T\psi _{\alpha }\left( X\right) \psi _{\beta }\left( X^{\prime
}\right) ...\right\rangle _{u} &=&\frac{1}{Z}\frac{\delta Z}{\left[ -\delta
J_{\alpha }^{+}\left( X\right) \right] \left[ -\delta J_{\beta }^{+}\left(
X^{\prime }\right) \right] ...}\text{,}  \notag
\end{eqnarray}%
where $X=\left( \mathbf{r,}\tau \right) $, $X^{\prime }=\left( \mathbf{r}%
^{\prime },\tau ^{\prime }\right) $. The correlators is (Matsubara) time
ordered. The generating functional\ for the connected correlations are $%
W=-\ln Z\left( J^{+},J\right) $:
\begin{eqnarray}
\left\langle \psi _{\alpha }\left( X\right) \right\rangle  &=&\frac{\delta W%
}{\delta J_{\alpha }^{+}\left( X\right) }\text{,}  \label{A3} \\
\left\langle T\psi _{\alpha }\left( X\right) \psi _{\beta }\left( X^{\prime
}\right) ...\right\rangle  &=&-\frac{\delta W}{\left[ -\delta J_{\alpha
}^{+}\left( X\right) \right] \left[ -\delta J_{\beta }^{+}\left( X^{\prime
}\right) \right] ...}\text{.}  \notag
\end{eqnarray}%
In particular, we define the normal and anomalous Green's functions (see ref.%
\cite{AGD}) as,%
\begin{eqnarray}
\left\langle T\psi _{\alpha }\left( X\right) \psi _{\beta }^{+}\left(
X^{\prime }\right) \right\rangle  &=&-\frac{\delta W}{\delta J_{\alpha
}\left( X\right) \delta J_{\beta }^{+}\left( X^{\prime }\right) }=-G_{\alpha
\beta }\left( X\mathbf{;}X^{\prime }\right) \text{;}  \notag \\
\left\langle T\psi _{\alpha }\left( X\right) \psi _{\beta }\left( X^{\prime
}\right) \right\rangle  &=&-\frac{\delta W}{\delta J_{\alpha }^{+}\left(
X\right) \delta J_{\beta }^{+}\left( X^{\prime }\right) }=F_{\alpha \beta
}\left( X\mathbf{;}X^{\prime }\right) \text{;}  \notag \\
\left\langle T\psi _{\alpha }^{+}\left( X\right) \psi _{\beta }^{+}\left(
X^{\prime }\right) \right\rangle  &=&-\frac{\delta W}{\delta J_{\alpha
}\left( X\right) \delta J_{\beta }\left( X^{\prime }\right) }=F_{\alpha
\beta }^{+}\left( X\mathbf{;}X^{\prime }\right)   \label{A4}
\end{eqnarray}%
where we denote $\left\langle \psi _{\alpha }\left( X\right) \right\rangle $
as $\psi _{\alpha }\left( X\right) $, and drop the time ordering operation $T
$ in correlators. For example, $\left\langle \psi _{\alpha }\left( X\right)
\psi _{\beta }\left( X^{\prime }\right) \right\rangle $ stands for $%
\left\langle T\psi _{\alpha }\left( X\right) \psi _{\beta }\left( X^{\prime
}\right) \right\rangle $.

\subsection{Derivation of the Gor'kov equation for superconductivity}

Using identities,
\begin{eqnarray}
\int D\psi ^{+}D\psi \frac{\delta }{\delta \psi _{\alpha }^{+}\left(
X\right) }e^{-\frac{1}{\hbar }S\left[ \psi ^{+}\mathbf{,}\psi ,J^{+},J\right]
} &=&0\text{;}  \label{A5} \\
\int D\psi ^{+}D\psi \frac{\delta }{\delta \psi _{\alpha }\left( X\right) }%
e^{-\frac{1}{\hbar }S\left[ \psi ^{+}\mathbf{,}\psi ,J^{+},J\right] } &=&0%
\text{,}  \notag
\end{eqnarray}%
the equations of motions lead to
\begin{gather}
\left( \hbar \partial _{\tau }\delta _{\alpha \beta }+K_{\alpha \beta
}\left( \mathbf{\nabla }\right) \right) \psi _{\beta }\left( X\right)
\mathbf{-}g\times   \notag \\
\left\langle \psi _{\beta }^{+}\left( X\right) \psi _{\beta }\left( X\right)
\psi _{\alpha }\left( X\right) \right\rangle =J_{\alpha }\left( \tau ,%
\mathbf{r}\right) \text{,}  \label{A6} \\
\left( \hbar \partial _{\tau }\delta _{\alpha \beta }-\widehat{K}_{\beta
\alpha }\left( -\mathbf{\nabla }\right) \right) \psi _{\beta }^{+}\left(
X\right) \mathbf{+}g\times   \notag \\
\left\langle \psi _{\alpha }^{+}\left( X\right) \psi _{\beta }^{+}\left(
X\right) \psi _{\beta }\left( X\right) \right\rangle =J_{\alpha }^{+}\left(
\tau ,\mathbf{r}\right) \text{. }  \notag
\end{gather}%
Generally full correlations of the fields can be expressed via connected
correlators. For example,

\begin{gather}
\left\langle \psi _{\beta }^{+}\left( X\right) \psi _{\beta }\left( X\right)
\psi _{\alpha }\left( X\right) \right\rangle =\psi _{\beta }^{+}\left(
X\right) \psi _{\beta }\left( X\right) \psi _{\alpha }\left( X\right) -
\notag \\
\psi _{\alpha }\left( X\right) \left\langle \psi _{\beta }\left( X\right)
\psi _{\beta }^{+}\left( X\right) \right\rangle _{c}+\psi _{\beta }\left(
X\right) \times  \label{A7} \\
\left\langle \psi _{\alpha }\left( X\right) \psi _{\beta }^{+}\left(
X\right) \right\rangle _{c}+\left\langle \psi _{\beta }^{+}\left( X\right)
\right\rangle \left\langle \psi _{\beta }\left( X\right) \psi _{\alpha
}\left( X\right) \right\rangle  \notag \\
+\left\langle \psi _{\beta }^{+}\left( X\right) \psi _{\beta }\left(
X\right) \psi _{\alpha }\left( X\right) \right\rangle _{c}\text{.}  \notag
\end{gather}

For Gaussian mean field approximation, we omit higher order connected
correlations, like $\left\langle \psi _{\beta }^{+}\left( X\right) \psi
_{\beta }\left( X\right) \psi _{\alpha }\left( X\right) \right\rangle $,

\begin{gather}
J_{\alpha }\left( X\right) =\left( \hbar \partial _{\tau }\delta _{\alpha
\beta }+K_{\alpha \beta }\right) \psi _{\beta }\left( X\right) \mathbf{-}%
g\times  \label{A8} \\
\psi _{\beta }^{+}\left( X\right) \psi _{\beta }\left( X\right) \psi
_{\alpha }\left( X\right) -\psi _{\alpha }\left( X\right) \left\langle \psi
_{\beta }\left( X\right) \psi _{\beta }^{+}\left( X\right) \right\rangle _{c}
\notag \\
+\psi _{\beta }\left( X\right) \left\langle \psi _{\alpha }\left( X\right)
\psi _{\beta }^{+}\left( X\right) \right\rangle _{c}+\psi _{\beta
}^{+}\left( X\right) \left\langle \psi _{\beta }\left( X\right) \psi
_{\alpha }\left( X\right) \right\rangle _{c}  \notag
\end{gather}%
Performing functional derivative$\frac{\delta }{\delta J_{\gamma }\left(
X^{\prime }\right) }$ (by using the identity, $\frac{\delta }{\delta
J_{\gamma }\left( X^{\prime }\right) }\psi _{\alpha }\left( X\right)
=\left\langle \psi _{\alpha }\left( X\right) \psi _{\gamma }^{+}\left(
X^{\prime }\right) \right\rangle $, the correlators of odd number of
Grassmannians vanishing) and taking at the end $J_{\alpha }\left( X\right)
=0 $, one obtains:

\begin{gather}
\delta \left( X\mathbf{-}X^{\prime }\right) \delta _{\alpha \gamma }=\left(
\partial _{\tau }\delta _{\alpha \beta }+\widehat{K}_{\alpha \beta }\right)
\left\langle \psi _{\beta }\left( X\right) \psi _{\gamma }^{+}\left(
X^{\prime }\right) \right\rangle \mathbf{-}g  \notag \\
\times \left\{ -\left\langle \psi _{\alpha }\left( X\right) \psi _{\gamma
}^{+}\left( X^{\prime }\right) \right\rangle \left\langle \psi _{\beta
}\left( X\right) \psi _{\beta }^{+}\left( X\right) \right\rangle \right.
\label{A9} \\
+\left\langle \psi _{\beta }\left( X\right) \psi _{\gamma }^{+}\left(
X^{\prime }\right) \right\rangle \left\langle \psi _{\alpha }\left( X\right)
\psi _{\beta }^{+}\left( X\right) \right\rangle  \notag \\
+\left. \left\langle \psi _{\beta }^{+}\left( X\right) \psi _{\gamma
}^{+}\left( X^{\prime }\right) \right\rangle \left\langle \psi _{\beta
}\left( X\right) \psi _{\alpha }\left( X\right) \right\rangle \right\} \text{%
,}  \notag
\end{gather}

or

\begin{gather}
\delta \left( X-X^{\prime }\right) \delta _{\alpha \gamma }=\left( -\partial
_{\tau }\delta _{\alpha \beta }-\widehat{K}_{\alpha \beta }\right) G_{\beta
\gamma }\left( X,X^{\prime }\right) \mathbf{-}g  \notag \\
\times \left\{ -G_{_{\alpha }\gamma }\left( X,X^{\prime }\right) G_{\beta
\beta }\left( X,X\right) \right.  \label{A10} \\
+G_{\beta \gamma }\left( X,X^{\prime }\right) G_{\alpha \beta }\left(
X,X\right) _{c}+  \notag \\
\left. F_{\beta \gamma }^{+}\left( X,X^{\prime }\right) F_{\beta \alpha
}\left( X,X\right) \right\} \text{.}  \notag
\end{gather}

\bigskip Similarly the second equation of motion,

\begin{gather}
J_{\alpha }^{+}\left( \tau ,\mathbf{r}\right) =\left( \partial _{\tau
}\delta _{\alpha \beta }-\widehat{K}_{\beta \alpha }\left( -\mathbf{\nabla }%
\right) \right) \left\langle \psi _{\beta }^{+}\left( \tau ,\mathbf{r}%
\right) \right\rangle \mathbf{+}g  \notag \\
\left\{ \left\langle \psi _{\alpha }^{+}\left( \tau ,\mathbf{r}\right) \psi
_{\beta }^{+}\left( \tau ,\mathbf{r}\right) \right\rangle \left\langle \psi
_{\beta }\left( \tau ,\mathbf{r}\right) \right\rangle \right.  \label{A11} \\
-\left\langle \psi _{\alpha }^{+}\left( \tau ,\mathbf{r}\right)
\right\rangle \left\langle \psi _{\beta }\left( \tau ,\mathbf{r}\right) \psi
_{\beta }^{+}\left( \tau ,\mathbf{r}\right) \right\rangle  \notag \\
+\left. \left\langle \psi _{\beta }^{+}\left( \tau ,\mathbf{r}\right)
\right\rangle \left\langle \psi _{\beta }\left( \tau ,\mathbf{r}\right) \psi
_{\alpha }^{+}\left( \tau ,\mathbf{r}\right) \right\rangle \right\} \text{,}
\notag
\end{gather}

gives
\begin{gather}
\left( \partial _{\tau }\delta _{\alpha \beta }-\widehat{K}_{\beta \alpha
}\left( -\mathbf{\nabla }\right) \right) F_{\beta \gamma }^{+}\left(
X,X^{\prime }\right) +g\left\{ -F_{\alpha \beta }^{+}\left( X,X\right)
\right.  \notag \\
G_{\beta \gamma }\left( X,X^{\prime }\right) +F_{\alpha \gamma }^{+}\left(
X,X^{\prime }\right) G_{\beta \beta }\left( X,X\right) +  \label{A12} \\
\left. G_{\beta \alpha }\left( X,X\right) F_{\beta \gamma }^{+}\left(
X,X^{\prime }\right) \right\} =0\text{.}  \notag
\end{gather}

In the superconducting phase, if the spin rotation invariance is not broken
and chiral symmetry is preserved, $G_{\alpha \beta }\left( X,X\right)
=G_{\alpha \beta }^{c}=n_{c}\delta _{\alpha \beta }$. For the $N$ component
spinors $n_{c}=n/N$, where $n$ is the density of electrons. The quadratic
parts of Eqs.(\ref{A10},\ref{A12}) simplify:
\begin{gather}
\mathbf{-}g\left\{ -G_{\alpha \gamma }\left( X,X^{\prime }\right) G_{\beta
\beta }\left( X,X\right) +G_{\beta \gamma }\left( X,X^{\prime }\right)
G_{\alpha \beta }\left( X,X\right) \right\}  \notag \\
=\mathbf{-}g\left\{ -G_{\alpha \gamma }\left( X,X^{\prime }\right)
Nn_{c}+n_{c}G_{\alpha \gamma }\left( X,X^{\prime }\right) \right\}  \notag \\
=g\left( N-1\right) n_{c}G_{\alpha \gamma }\left( X,X^{\prime }\right) ;
\label{A13} \\
g\left\{ F_{\alpha \gamma }^{+}\left( X,X^{\prime }\right) G_{\beta \beta
}\left( X,X\right) +G_{\beta \alpha }\left( X,X\right) F_{\beta \gamma
}^{+}\left( X,X^{\prime }\right) \right\}  \notag \\
=-g\left( N-1\right) n_{c}F_{\alpha \gamma }^{+}\left( X,X^{\prime }\right)
\text{,}  \notag
\end{gather}%
and such terms can be absorbed to the chemical potential term with the
chemical potential replaced by the renormalized

\begin{equation}
\mu +g\left( N-1\right) n_{c}=\mu _{R}\text{.}  \label{A14}
\end{equation}%
Therefore we finally obtain the Gor'kov equations, given in Eqs.(\ref{GeqX}).

\subsection{Derivation of DS equation and renormalized chemical potential
for nonsuperconducting state.}

We need only to discuss the equation for $\frac{\delta J_{\alpha }}{\delta
\psi _{\beta }}$, as$\frac{\delta J_{\alpha }^{+}}{\delta \psi _{\beta }^{+}}
$ is just the complex conjugate of $\frac{\delta J_{\alpha }}{\delta \psi
_{\beta }}$ and $\frac{\delta J_{\alpha }^{+}}{\delta \psi _{\beta }}=\frac{%
\delta J_{\alpha }}{\delta \psi _{\beta }^{+}}=0$. We reorganize the
equation $\frac{\delta J_{\alpha }}{\delta \psi _{\beta }}$as
\begin{gather}
\frac{\delta J_{\alpha }}{\delta \psi _{\beta }}=\left( \partial _{\tau
}\delta _{\alpha \beta }+K_{\alpha \beta }\right) \delta \left( X-X^{\prime
}\right) -g\delta \left( X-X^{\prime }\right) \times  \notag \\
\left\{ \left\langle \psi _{\alpha }\left( X\right) \psi _{\beta }^{+}\left(
X\right) \right\rangle _{c}-\frac{1}{4}\delta _{\alpha \beta }\sum
\left\langle \psi _{\gamma }\left( X\right) \psi _{\gamma }^{+}\left(
X\right) \right\rangle _{c}\right\}  \notag \\
+g\delta _{\alpha \beta }\delta \left( X-X^{\prime }\right) \left\{
\left\langle \psi _{\beta }\left( X\right) \psi _{\beta }^{+}\left( X\right)
\right\rangle _{c}-\right.  \label{A15} \\
\left. \sum \left\langle \psi _{\gamma }\left( X\right) \psi _{\gamma
}^{+}\left( X\right) \right\rangle _{c}\right\} \text{,}  \notag
\end{gather}

The last two terms are proportional to $\delta _{\alpha \beta }$, and can be
absorbed to the chemical potential,
\begin{gather}
\mu +g\delta _{\alpha \beta }\left\{ \sum \left\langle \psi _{\gamma }\left(
X\right) \psi _{\gamma }^{+}\left( X\right) \right\rangle _{c}\right.
\label{A16} \\
-\left. \left\langle \psi _{\beta }\left( X\right) \psi _{\beta }^{+}\left(
X\right) \right\rangle _{c}\right\} =\mu _{R\text{.}}  \notag
\end{gather}

The gap equation can be recasted as

\begin{eqnarray}
G^{-1} &=&G_{0}^{-1}+g\delta \left( X-X^{\prime }\right) G^{\prime }\left(
X;X^{\prime }\right)  \notag \\
G_{0}^{-1} &=&\left( \partial _{\tau }\delta _{\alpha \beta }+K_{\alpha
\beta }\right) \delta \left( X-X^{\prime }\right)  \label{A17} \\
G_{\alpha \beta }^{\prime }\left( X;X^{\prime }\right) &=&-\left\{
\left\langle \psi _{\alpha }\left( X\right) \psi _{\beta }^{+}\left(
X^{\prime }\right) \right\rangle _{c}-\right.  \notag \\
&&\left. \frac{1}{4}\delta _{\alpha \beta }\sum \left\langle \psi _{\gamma
}\left( X\right) \psi _{\gamma }^{+}\left( X^{\prime }\right) \right\rangle
_{c}\right\} ,  \notag
\end{eqnarray}%
where $G_{\alpha \beta }^{\prime }$ is the traceless part of $G_{\alpha
\beta }$, and the chemical potential in $K_{\alpha \beta }$ is $\mu _{R\text{%
.}}$

\section{A formula for energy of the superconducting state}

\subsection{Gap equation in the Nambu notation}

A compact representation of the Gorkov equations for superconductors is the
Nambu notations%
\begin{eqnarray}
\mathcal{G}_{\alpha \beta }\left( X,X^{\prime }\right) &=&%
\begin{pmatrix}
\frac{\delta \psi _{\alpha }\left( X\right) }{\delta J_{\beta }\left(
X^{\prime }\right) } & \frac{\delta \psi _{\alpha }\left( X\right) }{\delta
J_{\beta }^{+}\left( X^{\prime }\right) } \\
\frac{\delta \psi _{\alpha }^{+}\left( X\right) }{\delta J_{\beta }\left(
X^{\prime }\right) } & \frac{\delta \psi _{\alpha }^{+}\left( X\right) }{%
\delta J_{\beta }^{+}\left( X^{\prime }\right) }%
\end{pmatrix}
\label{B1} \\
&=&%
\begin{pmatrix}
-G_{\alpha \beta }\left( x,\tau ;x^{\prime },\tau ^{\prime }\right) &
F_{\alpha \beta }\left( x,\tau ;x^{\prime },\tau ^{\prime }\right) \\
F_{\alpha \beta }^{+}\left( x,\tau ;x^{\prime },\tau ^{\prime }\right) &
G_{\beta \alpha }\left( x^{\prime },\tau ^{\prime };x,\tau \right)%
\end{pmatrix}
\notag
\end{eqnarray}

so that

\begin{equation}
\begin{pmatrix}
\frac{\delta J_{\alpha }}{\delta \psi _{\beta }} & \frac{\delta J_{\alpha }}{%
\delta \psi _{\beta }^{+}} \\
\frac{\delta J_{\alpha }^{+}}{\delta \psi _{\beta }} & \frac{\delta
J_{\alpha }^{+}}{\delta \psi _{\beta }^{+}}%
\end{pmatrix}%
=\mathcal{G}_{\alpha \beta }^{-1}\text{.}  \label{B2}
\end{equation}%
.The functional identity

\begin{gather}
\frac{\delta J_{\alpha }\left( X\right) }{\delta \psi _{\beta }}\frac{\delta
\psi _{\beta }}{\delta J_{\gamma }\left( X^{\prime }\right) }+\frac{\delta
J_{\alpha }\left( X\right) }{\delta \psi _{\beta }^{+}}\times  \label{B3} \\
\frac{\delta \psi _{\beta }^{+}}{\delta J_{\gamma }\left( X^{\prime }\right)
}=\delta _{\alpha \gamma }\left( X-X^{\prime }\right) \text{,}  \notag
\end{gather}%
where abbreviations
\begin{gather}
\frac{\delta J_{\alpha }\left( X\right) }{\delta \psi _{\beta }}\frac{\delta
\psi _{\beta }}{\delta J_{\gamma }\left( X^{\prime }\right) }\equiv
\sum_{\beta }\int dX^{\prime \prime }\frac{\delta J_{\alpha }\left( X\right)
}{\delta \psi _{\beta }\left( X^{\prime \prime }\right) }\times \\
\frac{\delta \psi _{\beta }\left( X^{\prime \prime }\right) }{\delta
J_{\gamma }\left( X^{\prime }\right) }\text{, \ \ }\delta _{\alpha \gamma
}\left( X-X^{\prime }\right) =\delta _{\alpha \gamma }\delta \left(
X-X^{\prime }\right) ,  \notag
\end{gather}%
are used. In the Nambu matrix form it reads

\begin{equation}
\begin{pmatrix}
\frac{\delta J_{\alpha }}{\delta \psi _{\beta }} & \frac{\delta J_{\alpha }}{%
\delta \psi _{\beta }^{+}} \\
\frac{\delta J_{\alpha }^{+}}{\delta \psi _{\beta }} & \frac{\delta
J_{\alpha }^{+}}{\delta \psi _{\beta }^{+}}%
\end{pmatrix}%
\begin{pmatrix}
\frac{\delta \psi _{\beta }}{\delta J_{\gamma }} & \frac{\delta \psi _{\beta
}}{\delta J_{\gamma }^{+}} \\
\frac{\delta \psi _{\beta }^{+}}{\delta J_{\gamma }} & \frac{\delta \psi
_{\beta }^{+}}{\delta J_{\gamma }^{+}}%
\end{pmatrix}%
=%
\begin{pmatrix}
\delta _{\alpha \gamma } & 0 \\
0 & \delta _{\alpha \gamma }%
\end{pmatrix}%
\text{.}  \label{B4}
\end{equation}

\bigskip The derivatives are:%
\begin{eqnarray}
\frac{\delta J_{\alpha }\left( X\right) }{\delta \psi _{\beta }\left(
X^{\prime }\right) } &=&\left( \partial _{\tau }\delta _{\alpha \beta
}+K_{\alpha \beta }\right) \delta \left( X-X^{\prime }\right) ;  \notag \\
\frac{\delta J_{\alpha }\left( X\right) }{\delta \psi _{\beta }^{+}\left(
X^{\prime }\right) } &=&g\delta \left( X-X^{\prime }\right) \left\langle
\psi _{\alpha }\left( \tau ,\mathbf{r}\right) \psi _{\beta }\left( \tau ,%
\mathbf{r}\right) \right\rangle _{c};  \notag \\
\frac{\delta J_{\alpha }^{+}\left( X\right) }{\delta \psi _{\beta }\left(
X^{\prime }\right) } &=&g\delta \left( X-X^{\prime }\right) \left\langle
\psi _{\alpha }^{+}\left( \tau ,\mathbf{r}\right) \psi _{\beta }^{+}\left(
\tau ,\mathbf{r}\right) \right\rangle ;  \label{B5} \\
\frac{\delta J_{\alpha }^{+}\left( X\right) }{\delta \psi _{\beta
}^{+}\left( X^{\prime }\right) } &=&\left( \partial _{\tau }\delta _{\alpha
\beta }-K_{\beta \alpha }\left( -\mathbf{\nabla }\right) \right) \delta
\left( X-X^{\prime }\right) .  \notag
\end{eqnarray}%
For the non-interacting model

\begin{eqnarray}
\mathcal{G}_{0}^{-1} &=&%
\begin{pmatrix}
\left( \mathcal{G}_{0}^{-1}\right) _{11} & 0 \\
0 & \left( \mathcal{G}_{0}^{-1}\right) _{22}%
\end{pmatrix}%
\text{,}  \notag \\
\left( \mathcal{G}_{0}^{-1}\right) _{11} &=&\left( \partial _{\tau }\delta
_{\alpha \beta }+K_{\alpha \beta }\left( \mathbf{\nabla }\right) \right)
\delta \left( x-x^{\prime },\tau -\tau ^{\prime }\right) ;  \label{B6} \\
\left( \mathcal{G}_{0}^{-1}\right) _{22} &=&\left( \partial _{\tau }\delta
_{\alpha \beta }-K_{\beta \alpha }\left( -\mathbf{\nabla }\right) \right)
\delta \left( x-x^{\prime },\tau -\tau ^{\prime }\right) ,  \notag
\end{eqnarray}%
and the gap equation can be cast in the Dyson form

\begin{eqnarray}
\mathcal{G}^{-1} &=&\mathcal{G}_{0}^{-1}+%
\begin{pmatrix}
0 & \Sigma _{12} \\
\Sigma _{21} & 0%
\end{pmatrix}%
\text{;}  \notag \\
\Sigma _{12} &=&g\delta \left( x-x^{\prime },\tau -\tau ^{\prime }\right)
\left\langle \psi _{\alpha }\left( \tau ,\mathbf{r}\right) \psi _{\beta
}\left( \tau ,\mathbf{r}\right) \right\rangle _{c}\text{;} \\
\Sigma _{21} &=&g\delta \left( x-x^{\prime },\tau -\tau ^{\prime }\right)
\left\langle \psi _{\alpha }^{+}\left( \tau ,\mathbf{r}\right) \psi _{\beta
}^{+}\left( \tau ,\mathbf{r}\right) \right\rangle .  \notag
\end{eqnarray}

\subsection{Derivation of the expression for energy density}

The free energy is

\begin{equation}
\Omega \lbrack \mathcal{G}]=-\frac{1}{\beta }\left\{ {\frac{1}{2}}\text{Tr}%
\{-\ln \mathcal{G}+[\mathcal{G}_{0}^{-1}\mathcal{G}-1]\}+\Phi \lbrack
\mathcal{G}]\ \right\} \text{,}  \label{B8}
\end{equation}%
where in Gaussian approximation, \qquad \qquad \qquad \qquad \qquad

\begin{equation}
\Phi \lbrack \mathcal{G}]=\frac{g}{2}\int_{\tau ,r}F_{\alpha \beta }\left(
X;X\right) F_{\beta \alpha }^{+}\left( X;X\right) .  \label{B9}
\end{equation}%
The gap equation can be also obtained by $\frac{\delta }{\delta \mathcal{G}}%
\Omega \lbrack \mathcal{G}]=0$.

The energy difference between the superconducting state and normal state can
be obtained by the differentiating of the grand canonical potential with
respect to the coupling constant:$\frac{d}{dg}\Omega \lbrack \mathcal{G}]$.
The green function $\mathcal{G}$ is dependent on $g$, but due to $\frac{%
\delta }{\delta \mathcal{G}}\Omega \left[ \mathcal{G},g\right] =0$, \
\begin{eqnarray}
\frac{d}{dg}\Omega \left[ \mathcal{G},g\right] &=&\frac{\partial }{\partial g%
}\Omega \left[ \mathcal{G},g\right]  \label{B10} \\
&=&-\frac{1}{2\beta }\int_{\tau ,r}F_{\alpha \beta }\left( X;X\right)
F_{\beta \alpha }^{+}\left( X;X\right) \text{.}  \notag
\end{eqnarray}%
For homogeneous state, $F_{\alpha \beta }\left( X;X\right) ,F_{\beta \alpha
}^{+}\left( X;X\right) $ are constant \ (not dependent on $x,\tau $), and we
introduce

\begin{eqnarray}
F_{\alpha \beta }\left( X;X\right) &=&\frac{1}{g}\Delta _{\alpha \beta
},F_{\alpha \beta }^{+}\left( X;X\right)  \label{B11} \\
&=&\frac{1}{g}\Delta _{\alpha \beta }^{+}=\frac{1}{g}\Delta _{\beta \alpha
}^{\ast }\text{,}  \notag
\end{eqnarray}%
we obtain the free energy formula \cite{Abrikosov}%
\begin{eqnarray}
\frac{d}{dg}\Omega \lbrack \mathcal{G]}\mathcal{=-} &&\frac{V}{2}\frac{1}{%
g^{2}}\text{Tr}\left( \Delta \Delta ^{+}\right) \rightarrow  \label{B12} \\
d\Omega \lbrack \mathcal{G]} &\mathcal{=}&\frac{V}{2}d\left( \frac{1}{g}%
\right) \text{Tr}\left( \Delta \Delta ^{+}\right) .  \notag
\end{eqnarray}

\section{Gap equation and free energy for chiral symmetry
breaking states}

\subsection{Derivation of the gap equation}

We will discuss the non-superconducting state with $G_{\alpha \beta }\left(
X,X\right) \neq n_{c}\delta _{\alpha \beta }$, $F_{\alpha \beta }\left(
X;X^{\prime }\right) =0$,\ which happens for example in the case of chiral
symmetry breaking (CSB) state. We need only to discuss the equation for $%
\frac{\delta J_{\alpha }}{\delta \psi _{\beta }}$, as$\frac{\delta J_{\alpha
}^{+}}{\delta \psi _{\beta }^{+}}$ is just the complex conjugate of $\frac{%
\delta J_{\alpha }}{\delta \psi _{\beta }}$ and $\frac{\delta J_{\alpha }^{+}%
}{\delta \psi _{\beta }}=\frac{\delta J_{\alpha }}{\delta \psi _{\beta }^{+}}%
=0$. We reorganize the equation $\frac{\delta J_{\alpha }}{\delta \psi
_{\beta }}$as
\begin{gather}
\frac{\delta J_{\alpha }}{\delta \psi _{\beta }}=\left( \partial _{\tau
}\delta _{\alpha \beta }+\widehat{K}_{\alpha \beta }\right) \delta \left(
X-X^{\prime }\right) -g\delta \left( X-X^{\prime }\right) \times  \notag \\
\left\{ \left\langle \psi _{\alpha }\left( X\right) \psi _{\beta }^{+}\left(
X\right) \right\rangle _{c}-\frac{1}{4}\delta _{\alpha \beta }\right. \times
\\
\left. \sum \left\langle \psi _{\gamma }\left( X\right) \psi _{\gamma
}^{+}\left( X\right) \right\rangle _{c}\right\} +g\delta _{\alpha \beta
}\delta \left( X-X^{\prime }\right) \times  \notag \\
\left\{ \left\langle \psi _{\beta }\left( X\right) \psi _{\beta }^{+}\left(
X\right) \right\rangle _{c}-\sum \left\langle \psi _{\gamma }\left( X\right)
\psi _{\gamma }^{+}\left( X\right) \right\rangle _{c}\right\} \text{,}
\notag
\end{gather}

The last two terms are proportional to $\delta _{\alpha \beta }$, and can be
absorbed to the chemical potential,
\begin{gather}
\mu +g\delta _{\alpha \beta }\left\{ \sum \left\langle \psi _{\gamma }\left(
X\right) \psi _{\gamma }^{+}\left( X\right) \right\rangle _{c}-\right. \\
\left. \left\langle \psi _{\beta }\left( X\right) \psi _{\beta }^{+}\left(
X\right) \right\rangle _{c}\right\} =\mu _{R\text{.}}  \notag
\end{gather}%
The gap equation can be recasted as

\begin{eqnarray}
G^{-1} &=&G_{0}^{-1}+g\delta \left( X-X^{\prime }\right) G^{\prime }\left(
X;X^{\prime }\right)  \notag \\
G_{0}^{-1} &=&\left( \partial _{\tau }\delta _{\alpha \beta }+\widehat{K}%
_{\alpha \beta }\right) \delta \delta \left( X-X^{\prime }\right)  \notag \\
G_{\alpha \beta }^{\prime }\left( X;X^{\prime }\right) &=&-\left\{
\left\langle \psi _{\alpha }\left( X\right) \psi _{\beta }^{+}\left(
X^{\prime }\right) \right\rangle _{c}-\right.  \label{B14} \\
&&\left. \frac{1}{4}\delta _{\alpha \beta }\sum \left\langle \psi _{\gamma
}\left( X\right) \psi _{\gamma }^{+}\left( X^{\prime }\right) \right\rangle
_{c}\right\} ,  \notag
\end{eqnarray}%
where $G_{\alpha \beta }^{\prime }$ is the traceless part of $G_{\alpha
\beta }$, and the chemical potential in $\widehat{K}_{\alpha \beta }$ is $%
\mu _{R\text{.}}$The free energy is now
\begin{eqnarray}
\Omega \lbrack \mathcal{G}] &=&-\left( \beta \right) ^{-1}\left\{ \text{Tr}%
\{-\ln G+[G_{0}^{-1}G-1]\}\right. \\
&&\left. +\Phi \lbrack G]\right\} \text{,}  \notag
\end{eqnarray}%
where
\begin{equation}
\Phi \lbrack \mathcal{G}]=-\frac{g}{2}\int_{\tau ,r}G_{\alpha \beta
}^{\prime }\left( x,\tau ;x,\tau \right) G_{\beta \alpha }^{\prime }\left(
x,\tau ;x,\tau \right) \text{.}
\end{equation}%
A similar free energy equation can be obtained,

\begin{eqnarray}
\frac{d}{dg}\Omega \left[ \mathcal{G},g\right] &=&\frac{\partial }{\partial g%
}\Omega \left[ \mathcal{G},g\right]  \label{chiralfree} \\
&=&\frac{1}{2\beta }\int_{\tau ,r}G_{\alpha \beta }^{\prime }\left( x,\tau
;x,\tau \right) G_{\beta \alpha }^{\prime }\left( x,\tau ;x,\tau \right) .
\notag
\end{eqnarray}

\subsection{Details of the calculation of the condensate and energy}

Using the integral

\begin{gather}
\int_{\omega }\frac{1}{m^{2}+\left( \hbar v_{F}k\right) ^{2}-\left( \mu +i\
\omega \right) ^{2}\ }= \\
\frac{\pi }{\sqrt{m^{2}+\left( \hbar v_{F}k\right) ^{2}}}\Theta \left( \sqrt{%
m^{2}+\left( \hbar v_{F}k\right) ^{2}}-\mu \right) ,  \notag
\end{gather}

and
\begin{gather}
\frac{1}{\left( 2\pi \right) ^{3}}\int_{\omega ,p}\frac{1}{%
m^{2}+p^{2}-\left( \mu +i\ \omega \right) ^{2}\ }  \notag \\
=\frac{1}{4\pi }\int_{p=0}^{\Lambda }\frac{p\Theta \left( \sqrt{m^{2}+p^{2}}%
-\mu \right) }{\sqrt{m^{2}+p^{2}}} \\
=\left\{
\begin{array}{c}
\frac{1}{4\pi }\left( \sqrt{m^{2}+\Lambda ^{2}}-m\right) ,\mu \leq m \\
\frac{1}{4\pi }\left( \sqrt{m^{2}+\Lambda ^{2}}-\mu \right) ,\mu >m%
\end{array}%
\right.  \notag
\end{gather}

the gap equation becomes:
\begin{equation}
m=-\frac{g}{4\pi \left( \hbar v_{F}\right) ^{2}}\left\{
\begin{array}{c}
\left( \sqrt{m^{2}+\Lambda ^{2}}-m\right) ,\mu \leq m \\
\left( \sqrt{m^{2}+\Lambda ^{2}}-\mu \right) ,\mu >m%
\end{array}%
\right. \text{.}
\end{equation}

\section{Symmetries of the Dirac model}

\subsection{The chiral nonrelativistic $SU\left( 2\right) .$}

Generally the charge algebra is $\left[ Q,Q_{i}\right] =0;\left[ Q_{i},Q_{j}%
\right] =i\varepsilon _{ijk}Q_{k}$. Indeed all three chiral charges commute
with density:%
\begin{eqnarray}
\left[ \rho \left( x\right) ,Q_{i}\right] &=&\int_{y}\left[ \psi ^{\dagger
}\left( x\right) \psi \left( x\right) ,\psi ^{\dagger }\left( y\right)
T_{i}\psi \left( y\right) \right]  \label{C1a} \\
&=&\psi ^{\dagger }\left( x\right) \left[ I,T_{i}\right] \psi \left(
x\right) =0,  \notag
\end{eqnarray}%
and the kinetic term,

\begin{eqnarray}
\left[ H,Q_{i}\right] &=&\int_{y}\left[ \psi ^{\dagger }\left( x\right)
\left( \alpha _{1}p_{1}+\alpha _{2}p_{2}\right) \psi \left( x\right) ,\psi
^{\dagger }\left( y\right) T_{i}\psi \left( y\right) \right]  \notag \\
&=&\psi ^{\dagger }\left( x\right) \left[ \alpha _{1}p_{1}+\alpha
_{2}p_{2},T_{i}\right] \psi \left( x\right) =0.  \label{C2}
\end{eqnarray}

The three generator matrices, Eq.(\ref{Tmat}), constituting the $SU\left(
2\right) $, $\left[ T_{i},T_{j}\right] =i\varepsilon _{ijk}T_{k}$, commute
with both $\gamma _{0}\gamma _{1}$ and $\gamma _{0}\gamma _{2}$ that appear
in noninteracting Hamiltonian, Eq.(\ref{Hamiltonian}). The density - density
interactions part of Hamiltonian also commute with the $SU\left( 2\right) $
charges since

\begin{eqnarray}
\left[ \rho \left( r\right) ,Q_{i}\right] &=&\int_{r^{\prime }}\left[ \psi
^{\dagger }\left( r\right) \psi \left( r\right) ,\psi ^{\dagger }\left(
r^{\prime }\right) T_{i}\psi \left( r^{\prime }\right) \right]  \label{C3a}
\\
&=&\psi ^{\dagger }\left( r\right) \left[ I,T_{i}\right] \psi \left(
r\right) =0.  \notag
\end{eqnarray}

Correspondingly the action Eq.(\ref{relativistic action}) is invariant under
$\delta \psi =iT_{i}\psi ;\ \delta \overline{\psi }=i\overline{\psi }\gamma
_{0}T_{i}\gamma _{0}$. Indeed both the energy term,

\begin{eqnarray}
\delta A_{0} &=&-\int \left[ i\overline{\psi }\gamma _{0}T_{i}\gamma
_{0}\gamma _{0}\partial _{0}\psi +\overline{\psi }^{s}\gamma _{0}\partial
_{0}iT_{i}\psi \right] \\
&=&-\int \left[ -i\overline{\psi }\gamma _{0}T_{i}\partial _{0}\psi ^{s}+%
\overline{\psi }^{s}\gamma _{0}\partial _{0}iT_{i}\psi ^{s}\right] =0,
\notag
\end{eqnarray}%
\ and the momentum terms,

\begin{eqnarray}
\delta A_{1} &=&-\int \left[ i\overline{\psi }\gamma _{0}T_{i}\gamma
_{0}\left( \mathbf{\gamma }\cdot \mathbf{\partial }\right) \psi ^{s}+%
\overline{\psi }\left( \gamma \cdot \partial \right) iT_{i}\psi \right]
\notag \\
&=&-i\int \overline{\psi }\left[ \gamma _{0}T_{i}\gamma _{0}\mathbf{\gamma }+%
\mathbf{\gamma }T_{i}\right] \cdot \mathbf{\partial }\psi =0,
\end{eqnarray}%
are invariant. The nonrelativistic interaction term%
\begin{eqnarray}
\delta A &=&v\int \left[ i\overline{\psi }\gamma _{0}T_{i}\gamma _{0}\gamma
_{0}\psi +\overline{\psi }\gamma _{0}iT_{i}\psi \right] \left( \overline{%
\psi }^{r}\gamma _{0}\psi ^{r}\right) \\
&=&iv\int \left[ -\overline{\psi }\gamma _{0}T_{i}\psi +\overline{\psi }%
\gamma _{0}T_{i}\psi \right] \left( \overline{\psi }^{r}\gamma _{0}\psi
^{r}\right) =0.  \notag
\end{eqnarray}

\end{document}